\documentclass[aps,showpacs,prb,reprint,superscriptaddress,longbibliography]{revtex4-1}

\usepackage[colorlinks=true,linkcolor=blue,citecolor=blue,urlcolor=blue]{hyperref}

\usepackage{hyperref}
\usepackage{setspace} 
\usepackage{graphicx}
\usepackage{amsmath}
\usepackage{color}
\usepackage{amsmath}
\usepackage{amssymb}
\usepackage{verbatim}
\usepackage{latexsym}
\usepackage{enumerate} 
\usepackage{bm} 
\usepackage[caption=false]{subfig}
\begin{document}

\title{Hexagonal structure of phase III of solid hydrogen}

\author{Bartomeu Monserrat} 
\email{bm418@cam.ac.uk} 
\affiliation{Department of Physics and Astronomy, Rutgers
University, Piscataway, New Jersey 08854-8019, USA}
\affiliation{TCM Group, Cavendish Laboratory,
University of Cambridge, J.\ J.\ Thomson Avenue, Cambridge CB3 0HE,
United Kingdom}

\author{Richard J.\ Needs}
\affiliation{TCM Group, Cavendish Laboratory, University of Cambridge,
  J.\ J.\ Thomson Avenue, Cambridge CB3 0HE, United Kingdom}

\author{Eugene Gregoryanz} 
\affiliation{School of Physics and Centre
  for Science at Extreme Conditions, University of Edinburgh,
  Edinburgh EH9 3JZ, United Kingdom} \affiliation{Key Laboratory of
  Materials Physics, Institute of Solid State Physics, Chinese Academy
  of Sciences, Hefei, 230031, China}

\author{Chris J.\ Pickard}
\affiliation{Department of Materials Science $\&$ Metallurgy,
  University of Cambridge, 27 Charles Babbage Road, Cambridge CB3 0FS,
  United Kingdom}
\affiliation{Advanced
  Institute for Materials Research, Tohoku University 2-1-1 Katahira,
  Aoba, Sendai, 980-8577, Japan}

\date{\today}

\begin{abstract}

  A hexagonal structure of solid molecular hydrogen with $P6_122$
  symmetry is calculated to be more stable below about 200 GPa than
  the monoclinic $C2/c$ structure identified previously as the best
  candidate for phase III. We find that the effects of nuclear quantum 
  and thermal vibrations play a central role in the stabilization of 
  $P6_122$. The $P6_122$ and $C2/c$ structures are
  very similar and their Raman and infra-red data are in good
  agreement with experiment.  However, our calculations show that the
  hexagonal $P6_122$ structure provides better agreement with the
  available x-ray diffraction data than the $C2/c$ structure at
  pressures below about 200 GPa.  We suggest that two phase-III-like
  structures may be formed at high pressures, hexagonal $P6_122$ below
  about 200 GPa and monoclinic $C2/c$ at higher pressures.


\end{abstract}


\maketitle

\vspace{1cm}

\section{Introduction}

Experimental and theoretical studies of hydrogen at high pressures
have progressed rapidly in recent years. On the experimental front,
improvements in diamond anvil cell techniques have enabled the
exploration of static pressures above $300$~GPa in hydrogen,
\cite{silvera_h_review_1980,Mao_Hemley_1994,ceperley_rmp_h_he} and
even higher pressures in other materials.
\cite{Dubrovinsky_osmium_750GPa_2015}  The solid molecular
crystalline phases I, II, III, and IV/IV$^{\prime}$ of hydrogen have
been extensively studied experimentally. However, only the structure
of the low temperature and pressure phase I is known with precision,
and it is found to be a quantum solid consisting of molecules with
angular momentum $L=0$ in a mixture of ortho and para states and 
arranged on a hexagonal close packed lattice.
\cite{silvera_h_review_1980,Mao_Hemley_1994,ceperley_rmp_h_he} 
At low temperatures an pressures above about $100$~GPa ($25$~GPa for 
deuterium), hydrogen enters phase II in which the molecular rotations
are restricted.~\cite{h_phase_II} The detailed structure of phase II is
unknown, although infra-red (IR) and Raman vibrational data,
\cite{Mazin_H_1997,Goncharov_vibrations_2001} and x-ray and neutron
diffraction data
\cite{Loubeyre_1996,Kawamura_D_Xray_phase_II_2002,Goncharenko_II_2005,h_xray_phase_iii}
impose constraints on it.  At about 160 GPa, phase II transforms into
the ordered molecular phase III.~\cite{h_phase_III} Phase III has a
single strong IR active vibron peak and a much larger IR activity than
phase II. \cite{Hemley_H_1994}  The phases IV and IV$^{\prime}$, which
are similar to eachother, become stable at high pressures and temperatures
above about $300$~K.~\cite{h_phase_IV_eremets,h_phase_IV_gregoryanz,h_phase_IVp_gregoryanz}
They exhibit a high-frequency vibron peak that is weakly dependent on
pressure and a strong Raman vibron peak at lower frequencies which
softens rapidly with applied pressure. More recently, a new phase V
of hydrogen has been observed in Raman experiments around room temperature
reaching pressures of $388$~GPa,~\cite{gregoryanz_nature_phase_V}
but the structure of this new phase also remains unknown. 


On the theoretical front, high-pressure structures of hydrogen have
been investigated extensively using {\it ab initio} molecular dynamics,
\cite{Kohanoff_MD_1997,
  Kohanoff_MD_1999,md_raman_ackland,md_ramand_ir_azadi,Chen_MD_DMC_2014} 
path-integral molecular dynamics, \cite{Kitamura_H_PIMD_2000, pimd_hydrogen} quantum
Monte Carlo methods, \cite{Ceperley_1987_H, ceperley_rmp_h_he,
  azadi_hydrogen_gw_gap,
  prl_dissociation_hydrogen, Chen_MD_DMC_2014, h_dissociation_morales,
  hydrogen_nature_communications,azadi_h_metal_dmc_static}, 
first-principles density functional theory (DFT) 
methods,\cite{Johnson_Aschroft_2000,
nature_physics_h,phase_iv_prb,phase_iv_prb_erratum,azadi_hybrid_functionals}
and many-body methods.\cite{azadi_hydrogen_gw_gap,qp_excitonic_hydrogen_gw_bse}
The recent widespread adoption of DFT structure searching techniques
has led to the discovery of high-pressure hydrogen structures that are
consistent with many of the experimental observations.  Using the
\textit{ab initio} random structure searching (AIRSS) method, we found
a hydrogen structure of $P2_1/c$ symmetry that is a plausible model
for phase II. \cite{Pickard_PSS_2009}  We also discovered a monoclinic
structure of $C2/c$ symmetry and $24$ atoms per primitive cell
(henceforth called $C2/c$-$24$), that provides a good match to the
experimental vibrational data for phase III, and is the
lowest-enthalpy phase found over the pressure range in which phase III
is observed, of 160 to above 300 GPa. \cite{nature_physics_h}
Energetically competitive ``mixed structures'' of $C2$, $Pbcn$ and
$Ibam$ symmetries were also found~\cite{nature_physics_h} that consist
of alternate layers of strongly and weakly bonded molecules, which
provide simple models for phases IV/IV$^{\prime}$.  We have developed
improved models for these phases, in particular the $Pc$ structure
with 48 atoms per primitive unit cell,
\cite{phase_iv_prb,phase_iv_prb_erratum} and also a slightly better
structure with $96$ atoms per cell (see Supplemental Material of
Ref.~\onlinecite{phase_iv_prb}).  Structure searching methods have found
widespread application beyond hydrogen, discovering many new
structures that were subsequently synthesized. For example, AIRSS has
been used to determine structures of silane,
\cite{PhysRevLett.97.045504} aluminum hydride,
\cite{Aluminum_hydride_2007} ammonia,
\cite{Ammonia_2008,Ammonia_2014} ammonia hydrates,
\cite{ADMII_2009} and xenon oxides~\cite{xenon_oxides} that were subsequently verified by experiments.

Candidate structures for phases II, \cite{Pickard_PSS_2009} III,
\cite{nature_physics_h} and IV/IV$^{\prime}$ \cite{phase_iv_prb} have
been determined by structure searching using AIRSS. These searches did
not use experimental input, but the resulting structures 
provide Raman and IR vibrational data in reasonable agreement with 
experiment. Despite this success, there are still discrepancies between
theory and experiment. In particular, there remains an outstanding question 
about the structure of phase III. Although the vibrational signatures of the 
monoclinic $C2/c$-$24$ structure agree well with the experimental data for 
phase III, there is an inconsistency between the experimental x-ray 
diffraction data for phase III, reported in Ref.~\onlinecite{h_xray_phase_iii}, 
and the simulated x-ray data for $C2/c$-$24$. The experimental x-ray data 
are consistent with a hexagonal space group, but \textit{not} with the
monoclinic space group of $C2/c$-$24$.~\cite{h_exp_review}

In this work we investigate this discrepancy using DFT methods.
\cite{Jones_review_DFT_2015} We find a new hexagonal structure of high 
pressure hydrogen of $P6_122$ symmetry, that is calculated to be more
stable than the $C2/c$-$24$ structure below pressures of about $200$~GPa,
once the effects of quantum and thermal motion are incorporated.
The Raman and IR spectra of $P6_122$ are in good agreement with those 
observed experimentally for phase III, and the hexagonal symmetry leads
to the best agreement of any known candidate structure with the x-ray 
diffraction data available. We propose $P6_122$ as a candidate structure
for phase III of solid hydrogen.

The rest of the paper is organized as follows. In Sec.~\ref{sec:airss}
we describe the structure searches and in Sec.~\ref{sec:free_energy} we
calculate the relative free energies of the most competitive
candidate structures. We then characterize the new $P6_122$ structure
in Sec.~\ref{sec:properties}, and propose it as the candidate structure
of phase III of solid hydrogen in Sec.~\ref{sec:phaseIII}. We draw our
conclusions in Sec.~\ref{sec:conclusions}.

\section{Structure searches} \label{sec:airss}

We used AIRSS to search for low enthalpy static-lattice structures of
solid hydrogen at high pressures.
In contrast to previous searches, we focused on structures containing
a number of atoms or molecules equal to a highly composite
number. Highly composite numbers are positive integers that have more
divisors than any smaller positive integer, and searches over
structures containing a highly composite number of atoms or molecules
explore structures containing several different numbers of formula units 
during each search. For each
structure, a physically reasonable volume and set of atomic positions
were selected at random. Although some searches were performed without
symmetry constraints, for most searches we imposed common space group
symmetries of molecular crystals, and in particular space groups
$P2_1/c$, $P2_12_12_1$, $Pca2_1$, $Pna2_1$, and $C2/c$.  We
constrained the minimum initial atomic separations using data from
preliminary short AIRSS runs, with different minimum separations at
each pressure. This helps to space out the atoms appropriately while
retaining a high degree of randomness.  The structures were then
relaxed until the forces on the atoms were small and the pressure took
the required value. This procedure was repeated many times, and a
total of $85,424$ structures were generated. 

The searches were performed using the {\sc castep}~\cite{CASTEP} DFT 
plane-wave pseudopotential code with ``on the fly'' ultrasoft
pseudopotentials~\cite{PhysRevB.41.7892} and the BLYP density functional,
\cite{blyp_exchange, blyp_correlation} which has been shown to provide
a good description of molecular hydrogen at high
pressures.~\cite{clay_benchmarking} 
We employed an energy cut-off of $230$~eV,
and $\mathbf{k}$-point grids of spacings $2\pi\times0.07$~\AA$^{-1}$
and $2\pi\times0.05$~\AA$^{-1}$.

AIRSS found a previously 
unknown hydrogen structure of hexagonal $P6_122$ symmetry which is
energetically competitive with monoclinic $C2/c$-$24$.  A primitive
unit cell of $P6_122$ contains $36$ atoms, which is a highly composite
number.  It appears that this system size had not been explored
previously in hydrogen.

\begin{figure}
\centering
\includegraphics[scale=0.42]{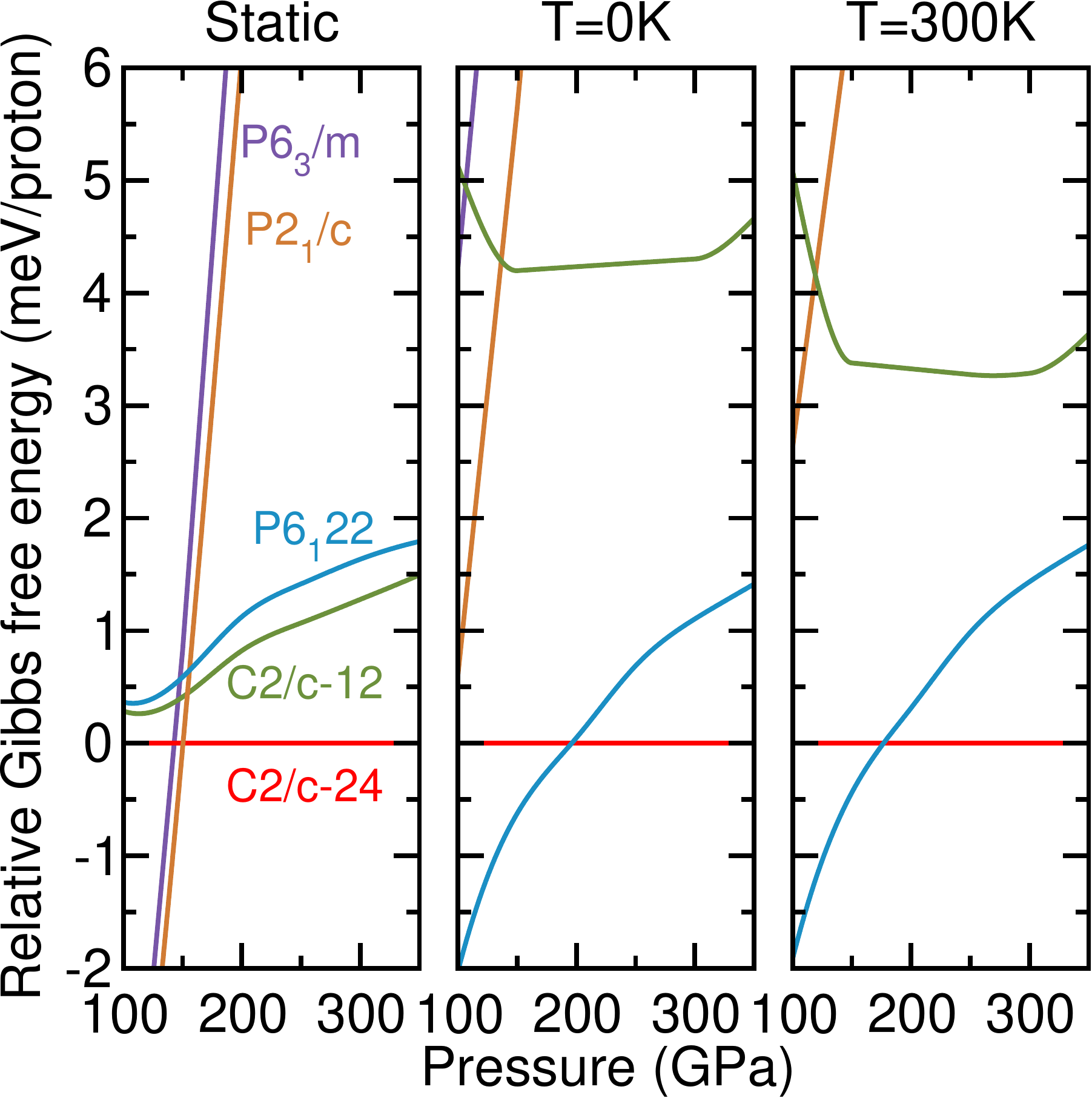}
\caption{Relative stability of $P6_122$ (blue), $C2/c$-$12$ (green),
  $P2_1/c$ (orange), and $P6_3/m$ (violet) with respect to $C2/c$-$24$
  (red) at the static lattice level, $T=0$~K, and $T=300$~K.}
\label{fig:energy_vs_pressure}
\end{figure}

\section{Free energy calculations} \label{sec:free_energy}

We next evaluate the relative enthalpies and free energies of the most
competitive structures of high pressure hydrogen in the pressure 
range $100$--$350$~GPa,
as shown in Fig.~\ref{fig:energy_vs_pressure}. We include the best
known candidate structures for phase II, of $P2_1/c$ and $P6_3/m$
symmetry, the best known candidates for phase III, the $C2/c$-$24$
structure, and a $12$ atoms variant, referred to as $C2/c$-$12$, as
well as the newly discovered $P6_122$ structure.

For both static lattice and vibrational energy calculations, 
we used an energy cut-off of
$1000$~eV and $\mathbf{k}$-point grids of spacing
$2\pi\times0.025$~\AA$^{-1}$. These parameters provide energy
differences between frozen-phonon structures that are converged to
better than $10^{-4}$~eV/atom, forces to better than $10^{-4}$~eV/\AA,
and stresses to better than $10^{-3}$~GPa.

The low mass of hydrogen leads to large vibrational energies and
amplitudes and to significant anharmonic nuclear motion, which must be
accounted for if accurate energies are to be calculated. We evaluated
the free energies using the method proposed in
Ref.~\onlinecite{PhysRevB.87.144302}.
The low-energy part of the Born-Oppenheimer energy surface was mapped
well beyond the harmonic region in a finite-displacement approach. We
took advantage of the recently introduced nondiagonal supercells
method~\cite{non_diagonal} to reach unprecedented levels of
convergence with respect to the size of the simulation cell. 
As an example, the
results reported here for the $P6_122$ structure were obtained using
nondiagonal supercells containing a maximum of $72$ atoms, but these
results are the same as those that would be obtained using the
standard supercell approach and a supercell containing $288$ atoms.
Tests with larger nondiagonal supercells containing a maximum of
$108$ atoms (equivalent to standard supercells containing $972$ atoms)
show that the final vibrational energies are converged to better than
$0.2$~meV/proton.  After construction of the anharmonic potential,
the resulting Schr\"{o}dinger equation was solved using a vibrational
self-consistent-field approach, in which the vibrational wave function
was represented in a basis of simple-harmonic-oscillator functions for
each degree of freedom, and converged results were achieved by
including up to $50$ basis functions per mode.

The relative enthalpies and free energies reported in 
Fig.~\ref{fig:energy_vs_pressure} correspond to 
static lattices, $T=0$~K, and $T=300$~K. At $300$~K,
$P6_122$ is thermodynamically stable at pressures below about
$180$~GPa when the vibrational energy is included.
The energy difference between $C2/c$-$24$ and $P6_122$ is small, but
it is clear that the new $P6_122$ structure is energetically
competitive in the pressure range where phase III is observed
experimentally.  The structural similarities between $P6_122$ and
$C2/c$-$24$ suggest that errors in the total free energies arising,
for example, from the choice of exchange-correlation functional,
should largely cancel when evaluating their relative free energies.
The $C2/c$-$12$ structure has a higher static lattice energy than
$C2/c$-$24$, and the inclusion of quantum and thermal vibrations
destabilizes it further.  This demonstrates the importance of the
stacking of layers in determining the relative stability of these
otherwise very similar structures. The candidate phase II structures
are significantly destabilized by the inclusion of quantum nuclear
motion, but it has recently been shown that a quantum Monte Carlo
description of the electronic energy is necessary to accurately
describe the relative energy of these structures compared to
$C2/c$-$24$.~\cite{hydrogen_nature_communications}  Finally, we note
that even at $300$~K, the vibrational energy is dominated by the
quantum zero-point motion.

We have also calculated the relative Gibbs free energy of
$P6_122$ with respect to $C2/c$-$24$ for the heavier deuterium
isotope. As atomic vibrations drive the thermodynamic stability of
$P6_122$ compared to $C2/c$-$24$, the heavier deuterium compound
has a narrower stability range. For example, at $300$~K the transition 
into the $C2/c$-$24$ structure is predicted to occur at about $160$~GPa.

\begin{figure} \centering
\subfloat[][(a) $P6_122$.]{
\includegraphics[scale=0.94]{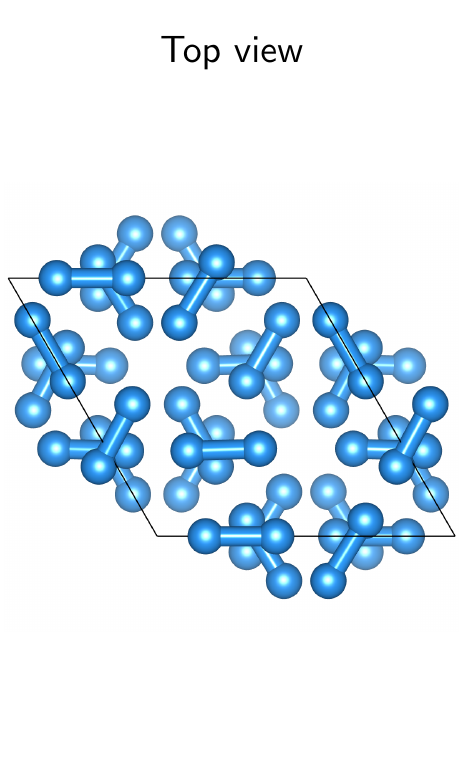}
\includegraphics[scale=0.94]{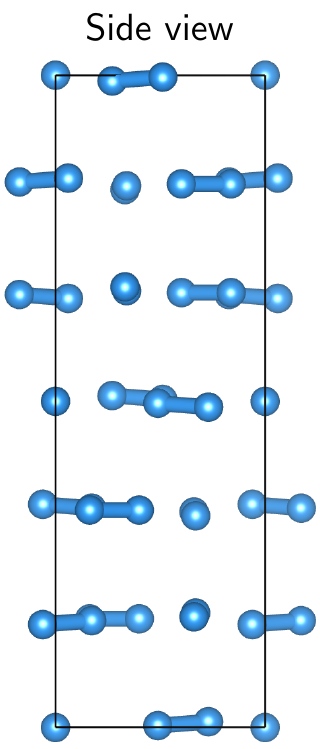}
\label{subfig:P6211}} \\
\subfloat[][(b) $C2/c$-$24$.]{
\includegraphics[scale=0.12]{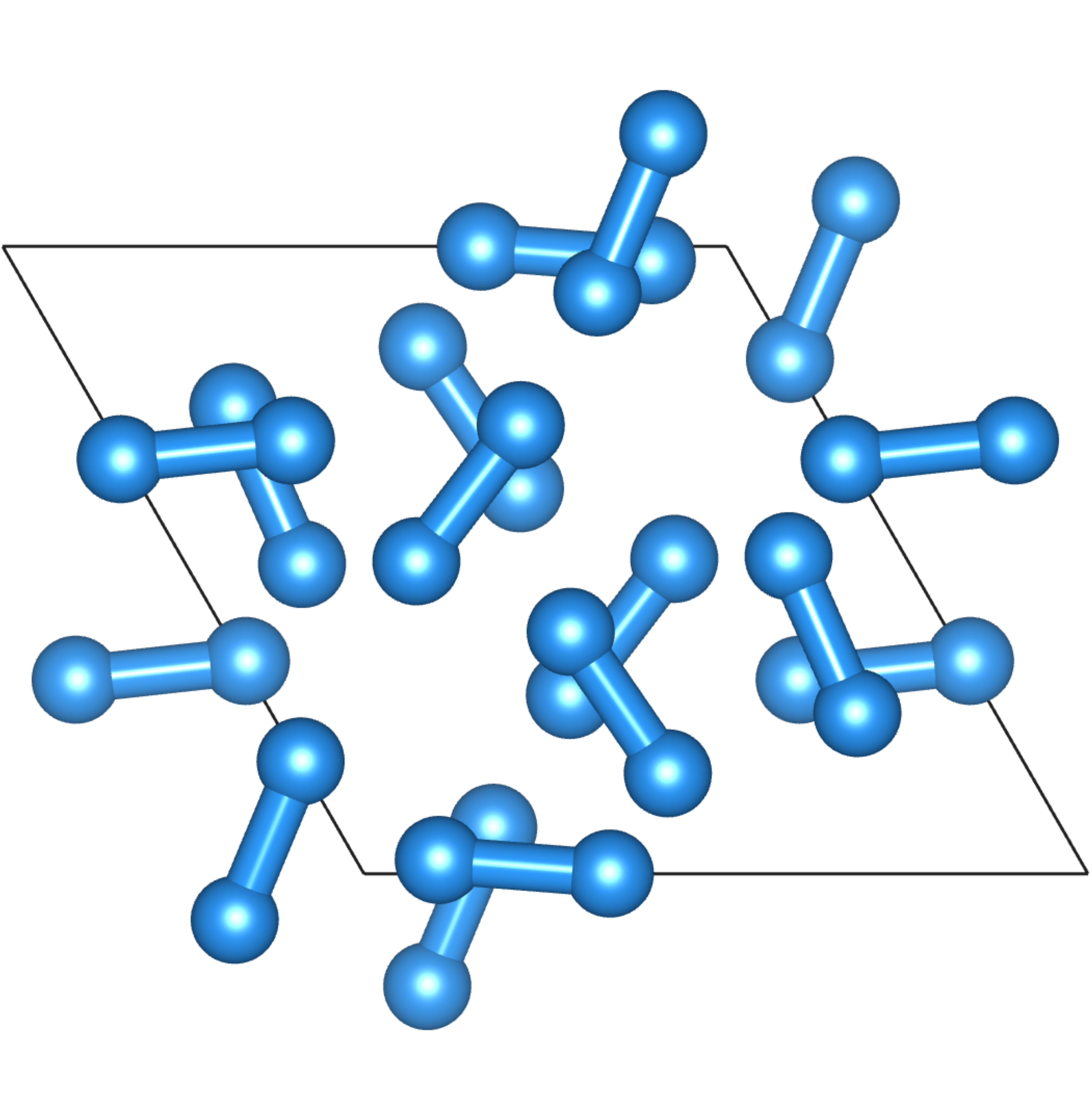}
\includegraphics[scale=0.12]{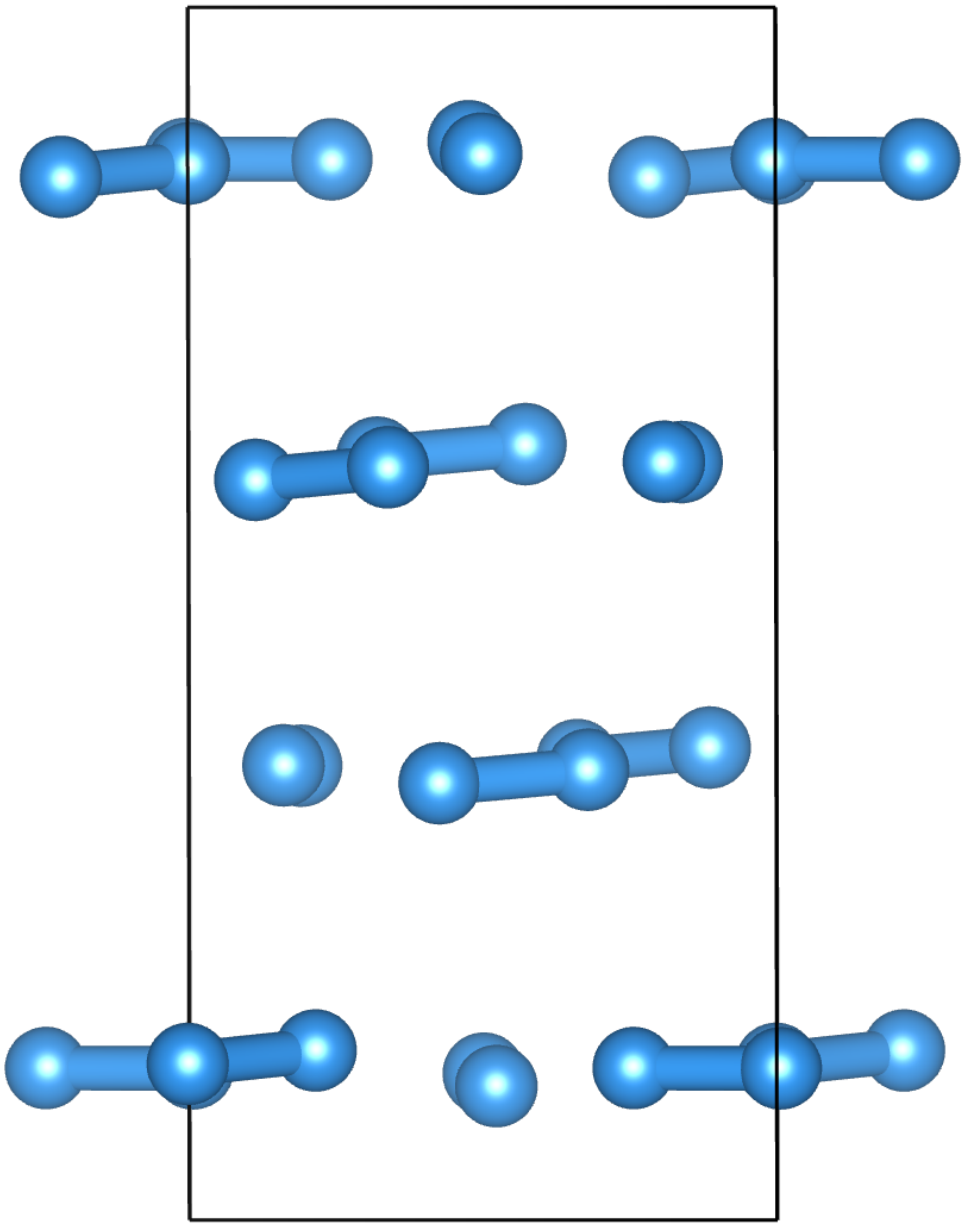}
\label{subfig:C2c}} \\
\caption{The $P6_122$ and $C2/c$-$24$ layered molecular structures.}
\label{fig:structures_C2/c-P6122}
\end{figure}


\begin{table*}
  \setlength{\tabcolsep}{7pt} 
  \caption{Static lattice structural details of $C2/c$-$24$ and $P6_122$ at $200$~GPa.}
  \label{tab:LatticeParams}
  \begin{ruledtabular}
  \begin{tabular}{c|ccccccccc}
            &  $a$   &   $b$  & $c$ & $\alpha$ & $\beta$ & $\gamma$ & Volume per proton & BL1 & BL2  \\ 
  \hline
  $C2/c$-$24$  &  $3.025$ \AA  & $3.025$ \AA & $5.408$ \AA & $90.1^\circ$ & $90.1^\circ$ & $119.9^\circ$ & $1.787$ \AA$^3$ & $0.719$ \AA & $0.716$ \AA \\
  $P6_122$  &  $3.022$ \AA  & $3.022$ \AA & $8.143$ \AA & $90.0^\circ$ & $90.0^\circ$ & $120.0^\circ$ & $1.789$ \AA$^3$ & $0.719$ \AA & $0.715$ \AA \\
\end{tabular}
\end{ruledtabular}
\end{table*}

\section{Properties of the $P6_122$ structure} \label{sec:properties}

The data reported in Fig.~\ref{fig:energy_vs_pressure} suggests that
$P6_122$ is a competitive candidate structure for phase III. Therefore,
in this section we characterize the $P6_122$ structure, and 
compare its spectroscopic signatures with experiment and
with those of $C2/c$-$24$, which is the best candidate for phase III
known at present.

\subsection{Structure}

Both $P6_122$ and $C2/c$-$24$ are layered molecular structures, with
two extra layers in the hexagonal primitive cell of $P6_122$,
as shown in Fig.~\ref{fig:structures_C2/c-P6122}. Their
structural details at $200$~GPa are provided in
Table~\ref{tab:LatticeParams}, which shows that their primitive cells
are similar, differing mainly in the length of the $c$ axis (about
$50$\% longer in $P6_122$ as a consequence of the two extra layers),
and in the slight monoclinic distortion in $C2/c$-$24$.  Two slightly
different molecular bond lengths (BL) appear in these structures, and
they differ by less than $0.001$~\AA\@ between the two structures.
The volume per proton of $P6_122$ is only $0.1$~\% larger than that 
of $C2/c$-$24$. We include a structure file of the $P6_122$ structure
as Supplemental Material.


\begin{figure}
\centering
\subfloat[][]{
\includegraphics[scale=0.40]{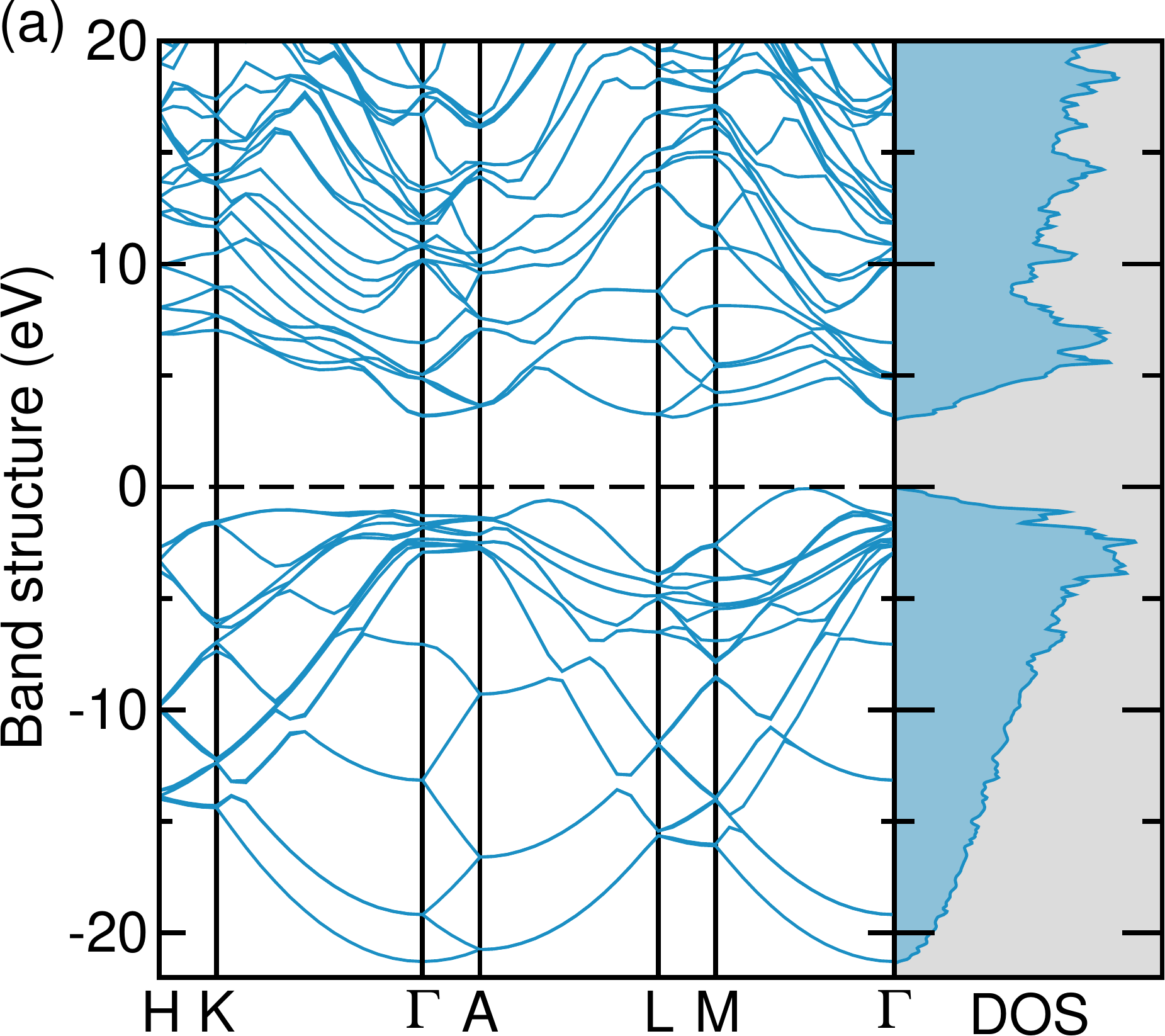}
\label{subfig:bs}
} \\
\subfloat[][]{
\includegraphics[scale=0.40]{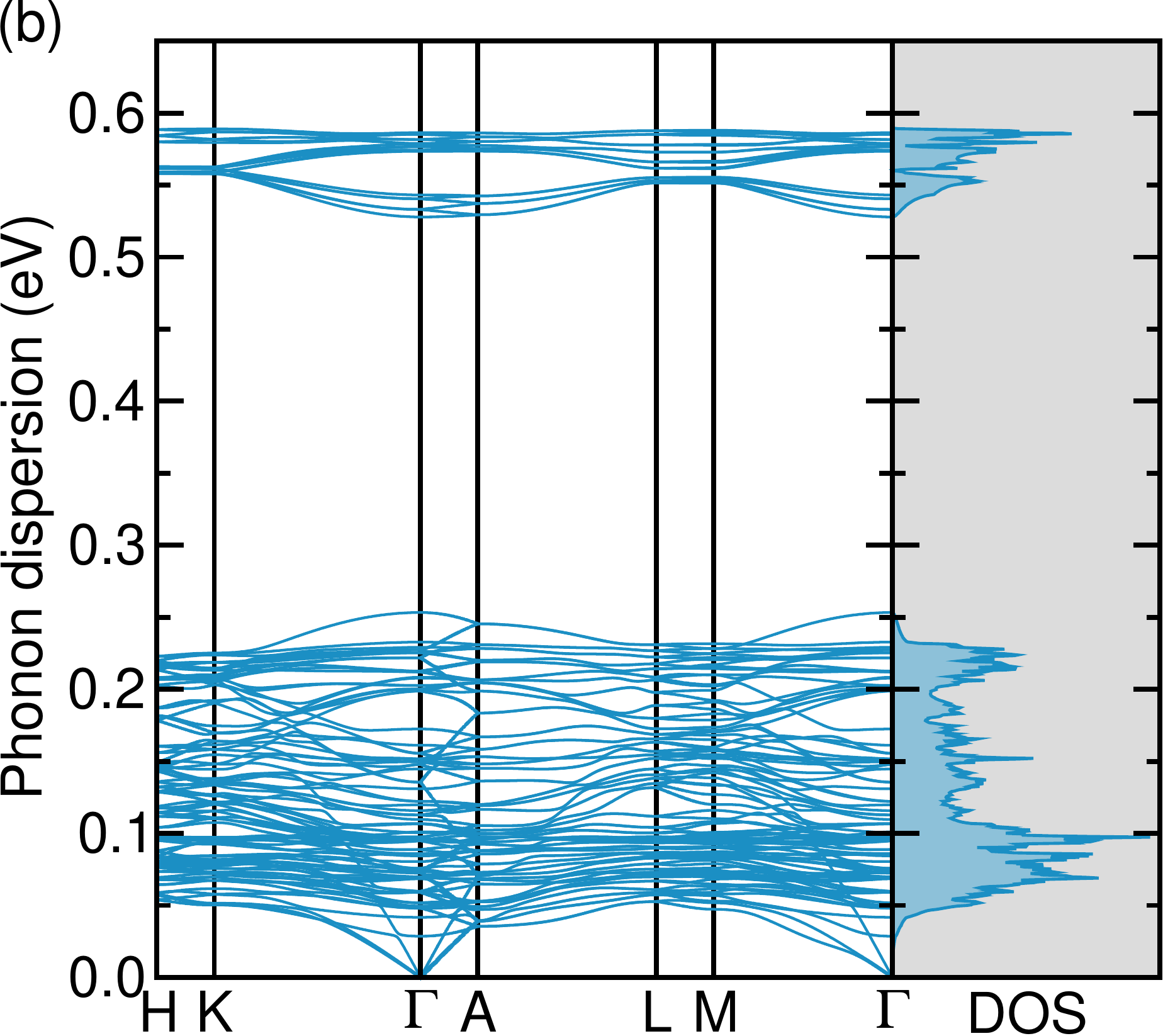}
\label{subfig:ph}
}
\caption{(a) Band structure along a high-symmetry path (left) and
  electronic density of states (right) of $P6_122$ at $200$~GPa. The
  dashed black line represents the Fermi level. (b) Phonon dispersion
  along a high-symmetry path (left) and vibrational density of states
  (right) of $P6_122$ at $200$~GPa.}
\label{fig:bs_ph}
\end{figure}

\subsection{Bandstructure and phonon dispersion}

In Fig.~\ref{subfig:bs} we show the band structure and density of
states of $P6_122$ at a pressure of $200$~GPa. In Fig.~\ref{subfig:ph}
we show the corresponding phonon dispersion and associated density of
vibrational states. The absence of imaginary frequencies in the phonon 
dispersion shows that $P6_122$ is a dynamically stable structure.

\begin{figure}
\centering
\includegraphics[scale=0.42]{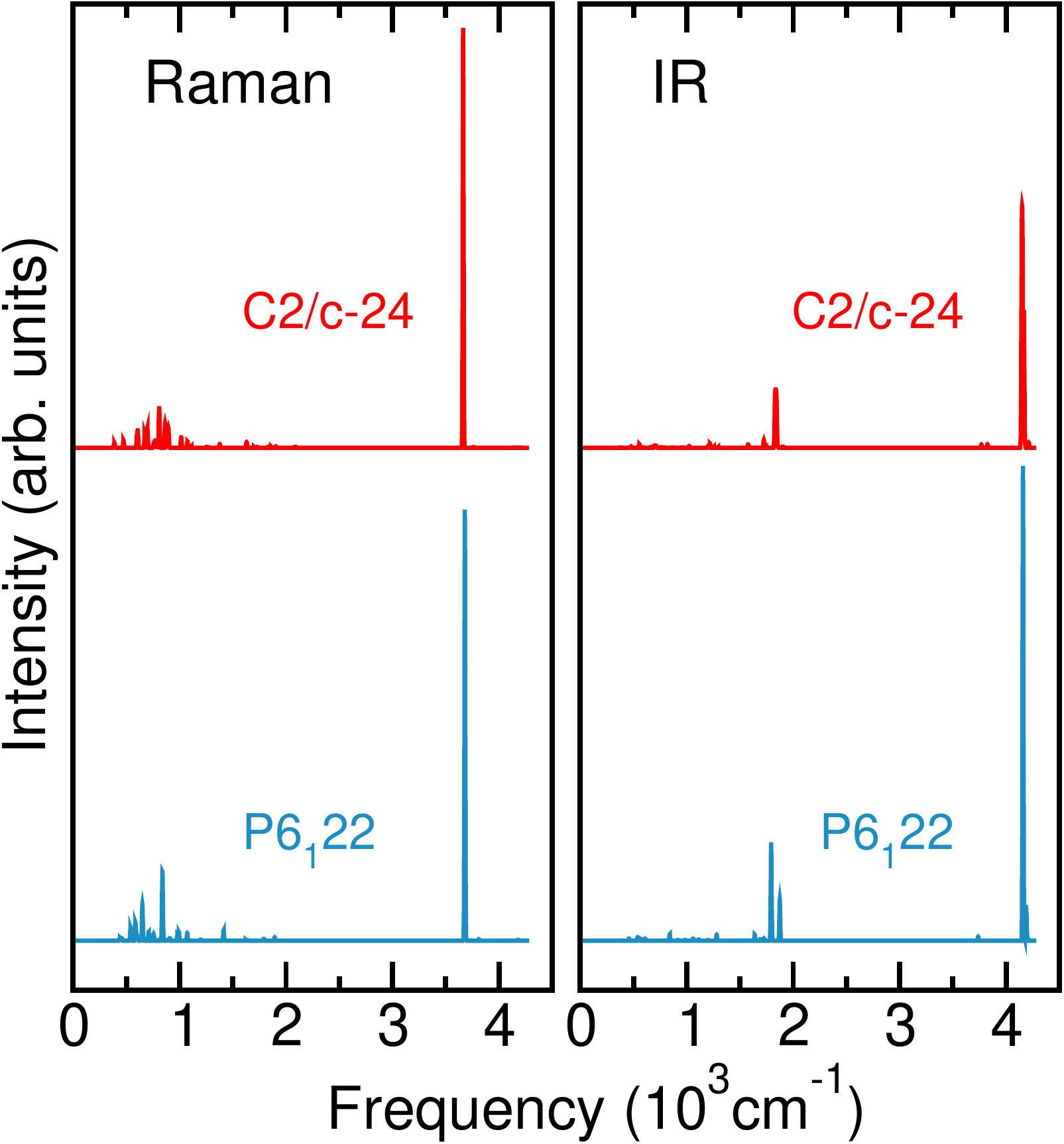}
\caption{Raman and IR spectra of $C2/c$-$24$ and $P6_122$ at
  $P=200$~GPa.}
\label{fig:spectra} 
\end{figure}

\subsection{Raman and IR}

The vast majority of experiments on pressurised hydrogen report Raman
and/or IR spectra. In Fig.~\ref{fig:spectra} we show the theoretical
Raman and IR spectra of $C2/c$-$24$ and $P6_122$ at $200$~GPa. As the
$C2/c$-$24$ and $P6_122$ structures are almost identical, the
frequencies of the active modes are indistinguishable, and agree well
with those observed experimentally.~\cite{Hemley_H_1994} The main
difference between the two signals is the stronger IR vibron peak for
$P6_122$, which is consistent with the observation that in phase III
the IR activity is much larger than in phase II.~\cite{Hemley_H_1994}
Overall, the IR and Raman spectra of $C2/c$-$24$ and $P6_122$ agree
well with the corresponding spectra observed for phase III, and
therefore we cannot unambiguously identify the structure of phase III
based purely on its vibrational response.

We note that the Raman and IR spectra were obtained
using the PBE functional~\cite{PhysRevLett.77.3865} instead of the
BLYP functional. The latter is not implemented in {\sc castep} within
the density functional perturbation theory formalism needed to
evaluate these spectra.  The main difference between the spectra
obtained using PBE and the one that would be obtained using a
different functional is the position of the peaks, caused by the
slightly different bond lengths predicted by the various functionals.

\begin{figure}
\centering
\includegraphics[scale=0.42]{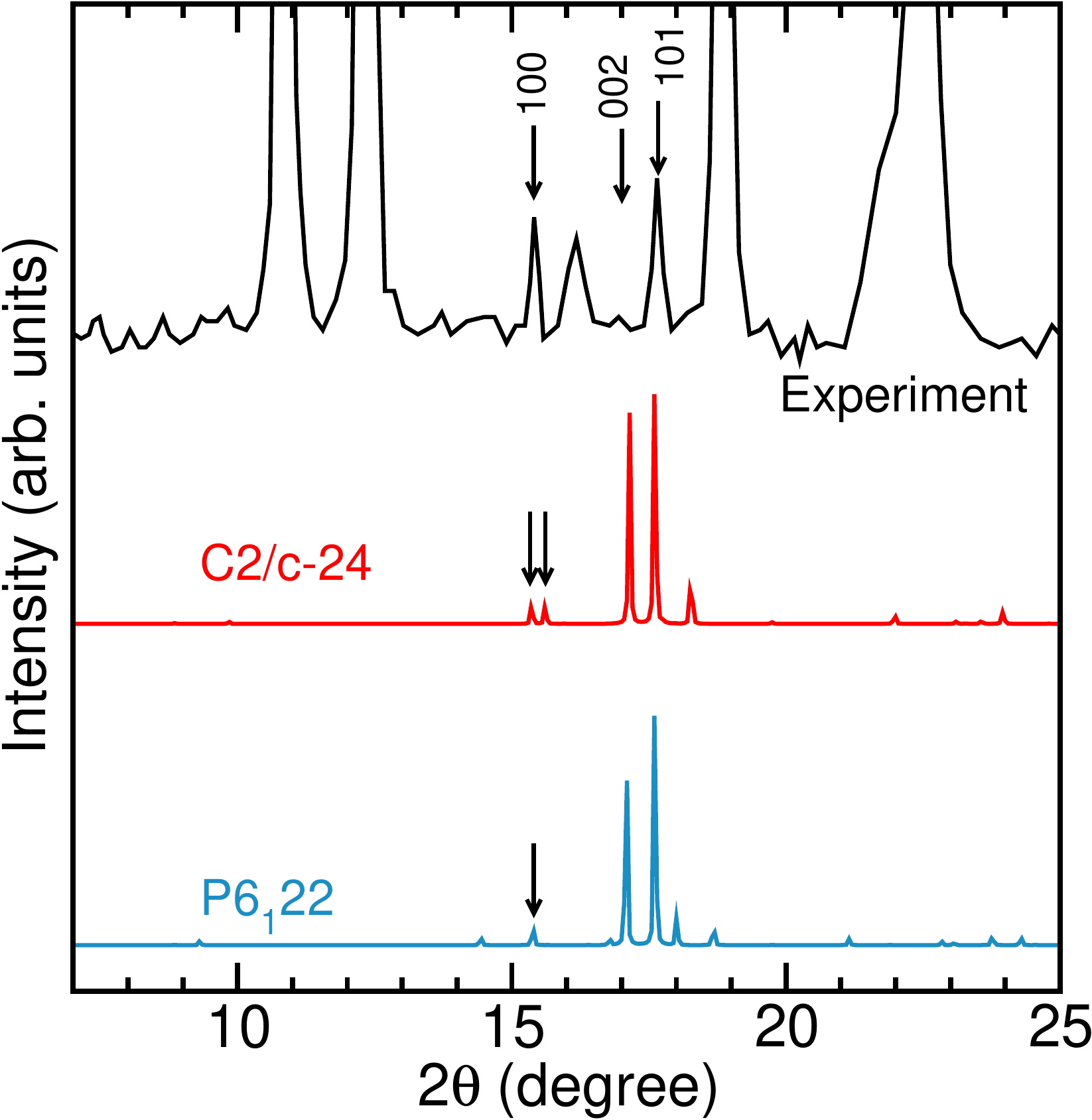}
\caption{Experimental x-ray diffraction data for phase
  III,~\cite{h_xray_phase_iii} and simulated x-ray diffraction data
  for the $C2/c$-$24$ and $P6_122$ structures at $P=174$~GPa and
  wavelength $\lambda=0.4122$~\AA. The black arrows in the
  experimental data indicate the H$_2$ peaks, while the other peaks
  correspond to the rhenium metal gasket. The two arrows in the calculated
  spectrum of $C2/c$-$24$ indicate the double peak at around
  $15.5^{\circ}$ caused by the monoclinic distortion,
  and the arrow in the $P6_122$ spectrum shows that in this 
  hexagonal structure there is a single peak around $15.5^{\circ}$.}
\label{fig:xray}
\end{figure}

\subsection{X-ray diffraction}

Hydrogen having the lowest atomic number Z is a very poor scatterer
of x-rays. This, combined with the restrictive access to a 
diamond anvil cell, makes the structural studies of hydrogen at 
high pressures notoriously difficult: even the structure of phase II,
which appears above $25$~GPa for deuterium, is still not known.
In a remarkable work, Akahama and co-workers
recently published x-ray data for phase III of hydrogen up to a
pressure of $183$~GPa.~\cite{h_xray_phase_iii}  Figure~\ref{fig:xray}
shows x-ray diffraction data from Ref.~\onlinecite{h_xray_phase_iii},
together with our data for $C2/c$-$24$ and $P6_122$, simulated using
the wavelength $\lambda=0.4122$~\AA\@ and the lattice parameters 
corresponding to a pressure of $174$~GPa.  The experimental data for phase
III 
shows two strong reflections at about $15.5^{\circ}$ and
$17.7^{\circ}$.  These data might show the continuation of the
$100$ and $101$ strong reflections observed in hexagonal phase I.
Experimentally, a weaker peak is also observed at about $17.1^{\circ}$
at some pressures, which could be the continuation of the $002$ peak
of phase I.  The $002$ peak is not easily observed because the crystallites
tend to grow with their $c$-axis perpendicular to the diamond culets,
and due to the geometrical constraints of the diamond anvil cell 
this prevents access to the $001$ planes. The available experimental data
from Ref.~\onlinecite{h_xray_phase_iii} suggests that 
the x-ray diffraction pattern of a good candidate structure for
phase III should (i) exhibit two peaks of similar intensity at
$15.5^{\circ}$ and at $17.7^{\circ}$, and (ii) could exhibit a weaker peak
at $17.1^{\circ}$.

The x-ray diffraction pattern of $C2/c$-$24$ has a double peak around
$15.5^{\circ}$ (see double arrow in Fig.~\ref{fig:xray}), resulting 
from the monoclinic distortion of the structure, and is inconsistent
with the number of peaks observed experimentally. This therefore, rules
out $C2/c$-$24$ as a possible structural candidate for phase III below 200 GPa.
$P6_122$ shows a single peak at $15.5^{\circ}$, which is consistent
with its hexagonal symmetry and with experiment.
However, the relative peak intensities of $P6_122$ are not in good agreement 
with
those observed experimentally. The strongest peaks in $P6_122$ are
those at $17.1^{\circ}$ and $17.7^{\circ}$, while the peak at
$15.5^{\circ}$ is weaker. The intensities of the calculated peaks would 
match experiment better if the peaks at $15.5^{\circ}$ and $17.1^{\circ}$ 
were interchanged.  

\section{Phase III of solid hydrogen} \label{sec:phaseIII}

The structural and vibrational characteristics
of $P6_122$, together with the energies reported in
Fig.~\ref{fig:energy_vs_pressure}, suggest that it is the best
candidate for phase III of solid hydrogen of all known structures. It is the first
hexagonal candidate for this phase, and its x-ray spectrum exhibits
the correct number of peaks at the appropriate locations.  The
remaining challenge is to explain the discrepancy in the intensities
of these peaks between theory and experiment.




Based on the data shown in Fig.~\ref{fig:energy_vs_pressure},
we speculate that phase III might in reality be two distinct phases,
with the $P6_122$ phase being stable below about $200$~GPa, and
$C2/c$-$24$ being stable at higher pressures. The almost identical
Raman and IR spectra of these two phases would make it difficult to
distinguish between them using spectroscopic techniques, and x-ray
data, which might distinguish between the hexagonal and monoclinic
symmetries, is only available up to $183$~GPa, where the hexagonal
structure is predicted to be stable.  It would be very useful if
experimental x-ray data could be collected above $200$~GPa.

\section{Conclusions} \label{sec:conclusions}

In conclusion, we have discovered a new candidate structure for phase
III of high pressure hydrogen with hexagonal symmetry and space
group $P6_122$. Our calculations suggest that $P6_122$ is the most stable
structure at pressures up to about $200$~GPa, and that its structural
and vibrational properties are in better agreement with experiment
than any other known candidate structures. Furthermore, the $C2/c$-$24$ 
structure is predicted to be stable at pressures above $200$~GPa, which
suggests that phase III might be two distinct phases, a hexagonal phase
below $200$~GPa, and a monoclinic phase at higher pressures.

\begin{acknowledgments}
  We thank Nicholas Worth for help with the x-ray diffraction pattern
  calculations.  B.M.\ acknowledges Robinson College, Cambridge, and
  the Cambridge Philosophical Society for a Henslow Research
  Fellowship. 
  R.J.N., E.G., and C.J.P. acknowledge financial support from the Engineering and Physical Sciences Research Council (EPSRC) of the United Kingdom (Grants No. EP/J017639/1, No. EP/J003999/1, and No. EP/K013688/1, respectively). C.J.P.\ is also supported by
  the Royal Society through a Royal Society Wolfson Research Merit
  award. 
  The calculations were performed on the Darwin Supercomputer
  of the University of Cambridge High Performance Computing Service
  facility (http://www.hpc.cam.ac.uk/) and 
the Archer facility of the UK national high performance computing service, for which access was obtained via the UKCP consortium and funded by EPSRC grant No. EP/K014560/1. 
\end{acknowledgments}

\bibliography{/Users/bartomeumonserrat/Documents/research/papers/references/anharmonic}

\begin{thebibliography}{53}%
\makeatletter
\providecommand \@ifxundefined [1]{%
 \@ifx{#1\undefined}
}%
\providecommand \@ifnum [1]{%
 \ifnum #1\expandafter \@firstoftwo
 \else \expandafter \@secondoftwo
 \fi
}%
\providecommand \@ifx [1]{%
 \ifx #1\expandafter \@firstoftwo
 \else \expandafter \@secondoftwo
 \fi
}%
\providecommand \natexlab [1]{#1}%
\providecommand \enquote  [1]{``#1''}%
\providecommand \bibnamefont  [1]{#1}%
\providecommand \bibfnamefont [1]{#1}%
\providecommand \citenamefont [1]{#1}%
\providecommand \href@noop [0]{\@secondoftwo}%
\providecommand \href [0]{\begingroup \@sanitize@url \@href}%
\providecommand \@href[1]{\@@startlink{#1}\@@href}%
\providecommand \@@href[1]{\endgroup#1\@@endlink}%
\providecommand \@sanitize@url [0]{\catcode `\\12\catcode `\$12\catcode
  `\&12\catcode `\#12\catcode `\^12\catcode `\_12\catcode `\%12\relax}%
\providecommand \@@startlink[1]{}%
\providecommand \@@endlink[0]{}%
\providecommand \url  [0]{\begingroup\@sanitize@url \@url }%
\providecommand \@url [1]{\endgroup\@href {#1}{\urlprefix }}%
\providecommand \urlprefix  [0]{URL }%
\providecommand \Eprint [0]{\href }%
\providecommand \doibase [0]{http://dx.doi.org/}%
\providecommand \selectlanguage [0]{\@gobble}%
\providecommand \bibinfo  [0]{\@secondoftwo}%
\providecommand \bibfield  [0]{\@secondoftwo}%
\providecommand \translation [1]{[#1]}%
\providecommand \BibitemOpen [0]{}%
\providecommand \bibitemStop [0]{}%
\providecommand \bibitemNoStop [0]{.\EOS\space}%
\providecommand \EOS [0]{\spacefactor3000\relax}%
\providecommand \BibitemShut  [1]{\csname bibitem#1\endcsname}%
\let\auto@bib@innerbib\@empty
\bibitem [{\citenamefont {Silvera}(1980)}]{silvera_h_review_1980}%
  \BibitemOpen
  \bibfield  {author} {\bibinfo {author} {\bibfnamefont {Isaac~F.}\
  \bibnamefont {Silvera}},\ }\bibfield  {title} {\enquote {\bibinfo {title}
  {The solid molecular hydrogens in the condensed phase: Fundamentals and
  static properties},}\ }\href
  {http://link.aps.org/doi/10.1103/RevModPhys.52.393} {\bibfield  {journal}
  {\bibinfo  {journal} {Rev. Mod. Phys.}\ }\textbf {\bibinfo {volume} {52}},\
  \bibinfo {pages} {393--452} (\bibinfo {year} {1980})}\BibitemShut {NoStop}%
\bibitem [{\citenamefont {Mao}\ and\ \citenamefont
  {Hemley}(1994)}]{Mao_Hemley_1994}%
  \BibitemOpen
  \bibfield  {author} {\bibinfo {author} {\bibfnamefont {Ho-kwang}\
  \bibnamefont {Mao}}\ and\ \bibinfo {author} {\bibfnamefont {Russell~J.}\
  \bibnamefont {Hemley}},\ }\bibfield  {title} {\enquote {\bibinfo {title}
  {Ultrahigh-pressure transitions in solid hydrogen},}\ }\href {\doibase
  10.1103/RevModPhys.66.671} {\bibfield  {journal} {\bibinfo  {journal} {Rev.
  Mod. Phys.}\ }\textbf {\bibinfo {volume} {66}},\ \bibinfo {pages} {671--692}
  (\bibinfo {year} {1994})}\BibitemShut {NoStop}%
\bibitem [{\citenamefont {McMahon}\ \emph {et~al.}(2012)\citenamefont
  {McMahon}, \citenamefont {Morales}, \citenamefont {Pierleoni},\ and\
  \citenamefont {Ceperley}}]{ceperley_rmp_h_he}%
  \BibitemOpen
  \bibfield  {author} {\bibinfo {author} {\bibfnamefont {Jeffrey~M.}\
  \bibnamefont {McMahon}}, \bibinfo {author} {\bibfnamefont {Miguel~A.}\
  \bibnamefont {Morales}}, \bibinfo {author} {\bibfnamefont {Carlo}\
  \bibnamefont {Pierleoni}}, \ and\ \bibinfo {author} {\bibfnamefont
  {David~M.}\ \bibnamefont {Ceperley}},\ }\bibfield  {title} {\enquote
  {\bibinfo {title} {The properties of hydrogen and helium under extreme
  conditions},}\ }\href {http://link.aps.org/doi/10.1103/RevModPhys.84.1607}
  {\bibfield  {journal} {\bibinfo  {journal} {Rev. Mod. Phys.}\ }\textbf
  {\bibinfo {volume} {84}},\ \bibinfo {pages} {1607--1653} (\bibinfo {year}
  {2012})}\BibitemShut {NoStop}%
\bibitem [{\citenamefont {Dubrovinsky}\ \emph {et~al.}(2015)\citenamefont
  {Dubrovinsky}, \citenamefont {Dubrovinskaia}, \citenamefont {Bykova},
  \citenamefont {Bykov}, \citenamefont {Prakapenka}, \citenamefont {Prescher},
  \citenamefont {Glazyrin}, \citenamefont {Liermann}, \citenamefont {Hanfland},
  \citenamefont {Ekholm}, \citenamefont {Feng}, \citenamefont {Pourovskii},
  \citenamefont {Katsnelson}, \citenamefont {Wills},\ and\ \citenamefont
  {Abrikosov}}]{Dubrovinsky_osmium_750GPa_2015}%
  \BibitemOpen
  \bibfield  {author} {\bibinfo {author} {\bibfnamefont {L.}~\bibnamefont
  {Dubrovinsky}}, \bibinfo {author} {\bibfnamefont {N.}~\bibnamefont
  {Dubrovinskaia}}, \bibinfo {author} {\bibfnamefont {E.}~\bibnamefont
  {Bykova}}, \bibinfo {author} {\bibfnamefont {M.}~\bibnamefont {Bykov}},
  \bibinfo {author} {\bibfnamefont {V.}~\bibnamefont {Prakapenka}}, \bibinfo
  {author} {\bibfnamefont {C.}~\bibnamefont {Prescher}}, \bibinfo {author}
  {\bibfnamefont {K.}~\bibnamefont {Glazyrin}}, \bibinfo {author}
  {\bibfnamefont {H.-P.}\ \bibnamefont {Liermann}}, \bibinfo {author}
  {\bibfnamefont {M.}~\bibnamefont {Hanfland}}, \bibinfo {author}
  {\bibfnamefont {M.}~\bibnamefont {Ekholm}}, \bibinfo {author} {\bibfnamefont
  {Q.}~\bibnamefont {Feng}}, \bibinfo {author} {\bibfnamefont {L.~V.}\
  \bibnamefont {Pourovskii}}, \bibinfo {author} {\bibfnamefont {M.~I.}\
  \bibnamefont {Katsnelson}}, \bibinfo {author} {\bibfnamefont {J.~M.}\
  \bibnamefont {Wills}}, \ and\ \bibinfo {author} {\bibfnamefont {I.~A.}\
  \bibnamefont {Abrikosov}},\ }\bibfield  {title} {\enquote {\bibinfo {title}
  {The most incompressible metal osmium at static pressures above 750
  gigapascals},}\ }\href {http://dx.doi.org/10.1038/nature14681} {\bibfield
  {journal} {\bibinfo  {journal} {Nature}\ }\textbf {\bibinfo {volume} {525}},\
  \bibinfo {pages} {226} (\bibinfo {year} {2015})}\BibitemShut {NoStop}%
\bibitem [{\citenamefont {Silvera}\ and\ \citenamefont
  {Wijngaarden}(1981)}]{h_phase_II}%
  \BibitemOpen
  \bibfield  {author} {\bibinfo {author} {\bibfnamefont {Isaac~F.}\
  \bibnamefont {Silvera}}\ and\ \bibinfo {author} {\bibfnamefont {Rinke~J.}\
  \bibnamefont {Wijngaarden}},\ }\bibfield  {title} {\enquote {\bibinfo {title}
  {New low-temperature phase of molecular deuterium at ultrahigh pressure},}\
  }\href {http://link.aps.org/doi/10.1103/PhysRevLett.47.39} {\bibfield
  {journal} {\bibinfo  {journal} {Phys. Rev. Lett.}\ }\textbf {\bibinfo
  {volume} {47}},\ \bibinfo {pages} {39--42} (\bibinfo {year}
  {1981})}\BibitemShut {NoStop}%
\bibitem [{\citenamefont {Mazin}\ \emph {et~al.}(1997)\citenamefont {Mazin},
  \citenamefont {Hemley}, \citenamefont {Goncharov}, \citenamefont {Hanfland},\
  and\ \citenamefont {Mao}}]{Mazin_H_1997}%
  \BibitemOpen
  \bibfield  {author} {\bibinfo {author} {\bibfnamefont {I.~I.}\ \bibnamefont
  {Mazin}}, \bibinfo {author} {\bibfnamefont {Russell~J.}\ \bibnamefont
  {Hemley}}, \bibinfo {author} {\bibfnamefont {A.~F.}\ \bibnamefont
  {Goncharov}}, \bibinfo {author} {\bibfnamefont {Michael}\ \bibnamefont
  {Hanfland}}, \ and\ \bibinfo {author} {\bibfnamefont {Ho-kwang}\ \bibnamefont
  {Mao}},\ }\bibfield  {title} {\enquote {\bibinfo {title} {Quantum and
  classical orientational ordering in solid hydrogen},}\ }\href
  {http://link.aps.org/doi/10.1103/PhysRevLett.78.1066} {\bibfield  {journal}
  {\bibinfo  {journal} {Phys. Rev. Lett.}\ }\textbf {\bibinfo {volume} {78}},\
  \bibinfo {pages} {1066--1069} (\bibinfo {year} {1997})}\BibitemShut {NoStop}%
\bibitem [{\citenamefont {Goncharov}\ \emph {et~al.}(2001)\citenamefont
  {Goncharov}, \citenamefont {Gregoryanz}, \citenamefont {Hemley},\ and\
  \citenamefont {Mao}}]{Goncharov_vibrations_2001}%
  \BibitemOpen
  \bibfield  {author} {\bibinfo {author} {\bibfnamefont {Alexander~F.}\
  \bibnamefont {Goncharov}}, \bibinfo {author} {\bibfnamefont {Eugene}\
  \bibnamefont {Gregoryanz}}, \bibinfo {author} {\bibfnamefont {Russell~J.}\
  \bibnamefont {Hemley}}, \ and\ \bibinfo {author} {\bibfnamefont {Ho-kwang}\
  \bibnamefont {Mao}},\ }\bibfield  {title} {\enquote {\bibinfo {title}
  {Spectroscopic studies of the vibrational and electronic properties of solid
  hydrogen to 285 {GP}a},}\ }\href
  {http://www.pnas.org/content/98/25/14234.abstract} {\bibfield  {journal}
  {\bibinfo  {journal} {Proc. Natl. Acad. Sci.}\ }\textbf {\bibinfo {volume}
  {98}},\ \bibinfo {pages} {14234--14237} (\bibinfo {year} {2001})}\BibitemShut
  {NoStop}%
\bibitem [{\citenamefont {Loubeyre}\ \emph {et~al.}(1996)\citenamefont
  {Loubeyre}, \citenamefont {LeToullec}, \citenamefont {Hausermann},
  \citenamefont {Hanfland}, \citenamefont {Hemley}, \citenamefont {Mao},\ and\
  \citenamefont {Finger}}]{Loubeyre_1996}%
  \BibitemOpen
  \bibfield  {author} {\bibinfo {author} {\bibfnamefont {P.}~\bibnamefont
  {Loubeyre}}, \bibinfo {author} {\bibfnamefont {R.}~\bibnamefont {LeToullec}},
  \bibinfo {author} {\bibfnamefont {D.}~\bibnamefont {Hausermann}}, \bibinfo
  {author} {\bibfnamefont {M.}~\bibnamefont {Hanfland}}, \bibinfo {author}
  {\bibfnamefont {R.~J.}\ \bibnamefont {Hemley}}, \bibinfo {author}
  {\bibfnamefont {H.~K.}\ \bibnamefont {Mao}}, \ and\ \bibinfo {author}
  {\bibfnamefont {L.~W.}\ \bibnamefont {Finger}},\ }\bibfield  {title}
  {\enquote {\bibinfo {title} {X-ray diffraction and equation of state of
  hydrogen at megabar pressures},}\ }\href {http://dx.doi.org/10.1038/383702a0}
  {\bibfield  {journal} {\bibinfo  {journal} {Nature}\ }\textbf {\bibinfo
  {volume} {383}},\ \bibinfo {pages} {702} (\bibinfo {year}
  {1996})}\BibitemShut {NoStop}%
\bibitem [{\citenamefont {Kawamura}\ \emph {et~al.}(2002)\citenamefont
  {Kawamura}, \citenamefont {Akahama}, \citenamefont {Umemoto}, \citenamefont
  {Takemura}, \citenamefont {Ohishi},\ and\ \citenamefont
  {Shimomura}}]{Kawamura_D_Xray_phase_II_2002}%
  \BibitemOpen
  \bibfield  {author} {\bibinfo {author} {\bibfnamefont {H.}~\bibnamefont
  {Kawamura}}, \bibinfo {author} {\bibfnamefont {Y.}~\bibnamefont {Akahama}},
  \bibinfo {author} {\bibfnamefont {S.}~\bibnamefont {Umemoto}}, \bibinfo
  {author} {\bibfnamefont {K.}~\bibnamefont {Takemura}}, \bibinfo {author}
  {\bibfnamefont {Y.}~\bibnamefont {Ohishi}}, \ and\ \bibinfo {author}
  {\bibfnamefont {O.}~\bibnamefont {Shimomura}},\ }\bibfield  {title} {\enquote
  {\bibinfo {title} {X-ray powder diffraction from solid deuterium},}\ }\href
  {http://stacks.iop.org/0953-8984/14/i=44/a=301} {\bibfield  {journal}
  {\bibinfo  {journal} {J. Phys.: Condens. Matter}\ }\textbf {\bibinfo {volume}
  {14}},\ \bibinfo {pages} {10407} (\bibinfo {year} {2002})}\BibitemShut
  {NoStop}%
\bibitem [{\citenamefont {Goncharenko}\ and\ \citenamefont
  {Loubeyre}(2005)}]{Goncharenko_II_2005}%
  \BibitemOpen
  \bibfield  {author} {\bibinfo {author} {\bibfnamefont {Igor}\ \bibnamefont
  {Goncharenko}}\ and\ \bibinfo {author} {\bibfnamefont {Paul}\ \bibnamefont
  {Loubeyre}},\ }\bibfield  {title} {\enquote {\bibinfo {title} {Neutron and
  x-ray diffraction study of the broken symmetry phase transition in solid
  deuterium},}\ }\href {http://dx.doi.org/10.1038/nature03699} {\bibfield
  {journal} {\bibinfo  {journal} {Nature}\ }\textbf {\bibinfo {volume} {435}},\
  \bibinfo {pages} {1206} (\bibinfo {year} {2005})}\BibitemShut {NoStop}%
\bibitem [{\citenamefont {Akahama}\ \emph {et~al.}(2010)\citenamefont
  {Akahama}, \citenamefont {Nishimura}, \citenamefont {Kawamura}, \citenamefont
  {Hirao}, \citenamefont {Ohishi},\ and\ \citenamefont
  {Takemura}}]{h_xray_phase_iii}%
  \BibitemOpen
  \bibfield  {author} {\bibinfo {author} {\bibfnamefont {Yuichi}\ \bibnamefont
  {Akahama}}, \bibinfo {author} {\bibfnamefont {Manabu}\ \bibnamefont
  {Nishimura}}, \bibinfo {author} {\bibfnamefont {Haruki}\ \bibnamefont
  {Kawamura}}, \bibinfo {author} {\bibfnamefont {Naohisa}\ \bibnamefont
  {Hirao}}, \bibinfo {author} {\bibfnamefont {Yasuo}\ \bibnamefont {Ohishi}}, \
  and\ \bibinfo {author} {\bibfnamefont {Kenichi}\ \bibnamefont {Takemura}},\
  }\bibfield  {title} {\enquote {\bibinfo {title} {Evidence from x-ray
  diffraction of orientational ordering in phase {III} of solid hydrogen at
  pressures up to 183 {GP}a},}\ }\href
  {http://link.aps.org/doi/10.1103/PhysRevB.82.060101} {\bibfield  {journal}
  {\bibinfo  {journal} {Phys. Rev. B}\ }\textbf {\bibinfo {volume} {82}},\
  \bibinfo {pages} {060101} (\bibinfo {year} {2010})}\BibitemShut {NoStop}%
\bibitem [{\citenamefont {Hemley}\ and\ \citenamefont
  {Mao}(1988)}]{h_phase_III}%
  \BibitemOpen
  \bibfield  {author} {\bibinfo {author} {\bibfnamefont {R.~J.}\ \bibnamefont
  {Hemley}}\ and\ \bibinfo {author} {\bibfnamefont {H.~K.}\ \bibnamefont
  {Mao}},\ }\bibfield  {title} {\enquote {\bibinfo {title} {Phase transition in
  solid molecular hydrogen at ultrahigh pressures},}\ }\href
  {http://link.aps.org/doi/10.1103/PhysRevLett.61.857} {\bibfield  {journal}
  {\bibinfo  {journal} {Phys. Rev. Lett.}\ }\textbf {\bibinfo {volume} {61}},\
  \bibinfo {pages} {857--860} (\bibinfo {year} {1988})}\BibitemShut {NoStop}%
\bibitem [{\citenamefont {Hemley}\ \emph {et~al.}(1994)\citenamefont {Hemley},
  \citenamefont {Soos}, \citenamefont {Hanfland},\ and\ \citenamefont
  {Mao}}]{Hemley_H_1994}%
  \BibitemOpen
  \bibfield  {author} {\bibinfo {author} {\bibfnamefont {Russell~J.}\
  \bibnamefont {Hemley}}, \bibinfo {author} {\bibfnamefont {Zoltan~G.}\
  \bibnamefont {Soos}}, \bibinfo {author} {\bibfnamefont {Michael}\
  \bibnamefont {Hanfland}}, \ and\ \bibinfo {author} {\bibfnamefont {Ho-kwang}\
  \bibnamefont {Mao}},\ }\bibfield  {title} {\enquote {\bibinfo {title}
  {Charge-transfer states in dense hydrogen},}\ }\href
  {http://dx.doi.org/10.1038/369384a0} {\bibfield  {journal} {\bibinfo
  {journal} {Nature}\ }\textbf {\bibinfo {volume} {369}},\ \bibinfo {pages}
  {384} (\bibinfo {year} {1994})}\BibitemShut {NoStop}%
\bibitem [{\citenamefont {Eremets}\ and\ \citenamefont
  {Troyan}(2011)}]{h_phase_IV_eremets}%
  \BibitemOpen
  \bibfield  {author} {\bibinfo {author} {\bibfnamefont {M.~I.}\ \bibnamefont
  {Eremets}}\ and\ \bibinfo {author} {\bibfnamefont {I.~A.}\ \bibnamefont
  {Troyan}},\ }\bibfield  {title} {\enquote {\bibinfo {title} {Conductive dense
  hydrogen},}\ }\href {http://dx.doi.org/10.1038/nmat3175} {\bibfield
  {journal} {\bibinfo  {journal} {Nat. Mater.}\ }\textbf {\bibinfo {volume}
  {10}},\ \bibinfo {pages} {927} (\bibinfo {year} {2011})}\BibitemShut
  {NoStop}%
\bibitem [{\citenamefont {Howie}\ \emph
  {et~al.}(2012{\natexlab{a}})\citenamefont {Howie}, \citenamefont {Guillaume},
  \citenamefont {Scheler}, \citenamefont {Goncharov},\ and\ \citenamefont
  {Gregoryanz}}]{h_phase_IV_gregoryanz}%
  \BibitemOpen
  \bibfield  {author} {\bibinfo {author} {\bibfnamefont {Ross~T.}\ \bibnamefont
  {Howie}}, \bibinfo {author} {\bibfnamefont {Christophe~L.}\ \bibnamefont
  {Guillaume}}, \bibinfo {author} {\bibfnamefont {Thomas}\ \bibnamefont
  {Scheler}}, \bibinfo {author} {\bibfnamefont {Alexander~F.}\ \bibnamefont
  {Goncharov}}, \ and\ \bibinfo {author} {\bibfnamefont {Eugene}\ \bibnamefont
  {Gregoryanz}},\ }\bibfield  {title} {\enquote {\bibinfo {title} {Mixed
  molecular and atomic phase of dense hydrogen},}\ }\href
  {http://link.aps.org/doi/10.1103/PhysRevLett.108.125501} {\bibfield
  {journal} {\bibinfo  {journal} {Phys. Rev. Lett.}\ }\textbf {\bibinfo
  {volume} {108}},\ \bibinfo {pages} {125501} (\bibinfo {year}
  {2012}{\natexlab{a}})}\BibitemShut {NoStop}%
\bibitem [{\citenamefont {Howie}\ \emph
  {et~al.}(2012{\natexlab{b}})\citenamefont {Howie}, \citenamefont {Scheler},
  \citenamefont {Guillaume},\ and\ \citenamefont
  {Gregoryanz}}]{h_phase_IVp_gregoryanz}%
  \BibitemOpen
  \bibfield  {author} {\bibinfo {author} {\bibfnamefont {Ross~T.}\ \bibnamefont
  {Howie}}, \bibinfo {author} {\bibfnamefont {Thomas}\ \bibnamefont {Scheler}},
  \bibinfo {author} {\bibfnamefont {Christophe~L.}\ \bibnamefont {Guillaume}},
  \ and\ \bibinfo {author} {\bibfnamefont {Eugene}\ \bibnamefont
  {Gregoryanz}},\ }\bibfield  {title} {\enquote {\bibinfo {title} {Proton
  tunneling in phase {IV} of hydrogen and deuterium},}\ }\href
  {http://link.aps.org/doi/10.1103/PhysRevB.86.214104} {\bibfield  {journal}
  {\bibinfo  {journal} {Phys. Rev. B}\ }\textbf {\bibinfo {volume} {86}},\
  \bibinfo {pages} {214104} (\bibinfo {year} {2012}{\natexlab{b}})}\BibitemShut
  {NoStop}%
\bibitem [{\citenamefont {Dalladay-Simpson}\ \emph {et~al.}(2016)\citenamefont
  {Dalladay-Simpson}, \citenamefont {Howie},\ and\ \citenamefont
  {Gregoryanz}}]{gregoryanz_nature_phase_V}%
  \BibitemOpen
  \bibfield  {author} {\bibinfo {author} {\bibfnamefont {Philip}\ \bibnamefont
  {Dalladay-Simpson}}, \bibinfo {author} {\bibfnamefont {Ross~T.}\ \bibnamefont
  {Howie}}, \ and\ \bibinfo {author} {\bibfnamefont {Eugene}\ \bibnamefont
  {Gregoryanz}},\ }\bibfield  {title} {\enquote {\bibinfo {title} {Evidence for
  a new phase of dense hydrogen above 325 gigapascals},}\ }\href
  {http://dx.doi.org/10.1038/nature16164} {\bibfield  {journal} {\bibinfo
  {journal} {Nature}\ }\textbf {\bibinfo {volume} {529}},\ \bibinfo {pages}
  {63} (\bibinfo {year} {2016})}\BibitemShut {NoStop}%
\bibitem [{\citenamefont {Kohanoff}\ \emph {et~al.}(1997)\citenamefont
  {Kohanoff}, \citenamefont {Scandolo}, \citenamefont {Chiarotti},\ and\
  \citenamefont {Tosatti}}]{Kohanoff_MD_1997}%
  \BibitemOpen
  \bibfield  {author} {\bibinfo {author} {\bibfnamefont {Jorge}\ \bibnamefont
  {Kohanoff}}, \bibinfo {author} {\bibfnamefont {Sandro}\ \bibnamefont
  {Scandolo}}, \bibinfo {author} {\bibfnamefont {Guido~L.}\ \bibnamefont
  {Chiarotti}}, \ and\ \bibinfo {author} {\bibfnamefont {Erio}\ \bibnamefont
  {Tosatti}},\ }\bibfield  {title} {\enquote {\bibinfo {title} {Solid molecular
  hydrogen: The broken symmetry phase},}\ }\href
  {http://link.aps.org/doi/10.1103/PhysRevLett.78.2783} {\bibfield  {journal}
  {\bibinfo  {journal} {Phys. Rev. Lett.}\ }\textbf {\bibinfo {volume} {78}},\
  \bibinfo {pages} {2783--2786} (\bibinfo {year} {1997})}\BibitemShut {NoStop}%
\bibitem [{\citenamefont {Kohanoff}\ \emph {et~al.}(1999)\citenamefont
  {Kohanoff}, \citenamefont {Scandolo}, \citenamefont {de~Gironcoli},\ and\
  \citenamefont {Tosatti}}]{Kohanoff_MD_1999}%
  \BibitemOpen
  \bibfield  {author} {\bibinfo {author} {\bibfnamefont {Jorge}\ \bibnamefont
  {Kohanoff}}, \bibinfo {author} {\bibfnamefont {Sandro}\ \bibnamefont
  {Scandolo}}, \bibinfo {author} {\bibfnamefont {Stefano}\ \bibnamefont
  {de~Gironcoli}}, \ and\ \bibinfo {author} {\bibfnamefont {Erio}\ \bibnamefont
  {Tosatti}},\ }\bibfield  {title} {\enquote {\bibinfo {title}
  {Dipole-quadrupole interactions and the nature of phase {III} of compressed
  hydrogen},}\ }\href {\doibase 10.1103/PhysRevLett.83.4097} {\bibfield
  {journal} {\bibinfo  {journal} {Phys. Rev. Lett.}\ }\textbf {\bibinfo
  {volume} {83}},\ \bibinfo {pages} {4097--4100} (\bibinfo {year}
  {1999})}\BibitemShut {NoStop}%
\bibitem [{\citenamefont {Magd\u{a}u}\ and\ \citenamefont
  {Ackland}(2013)}]{md_raman_ackland}%
  \BibitemOpen
  \bibfield  {author} {\bibinfo {author} {\bibfnamefont {Ioan~B.}\ \bibnamefont
  {Magd\u{a}u}}\ and\ \bibinfo {author} {\bibfnamefont {Graeme~J.}\
  \bibnamefont {Ackland}},\ }\bibfield  {title} {\enquote {\bibinfo {title}
  {Identification of high-pressure phases {III} and {IV} in hydrogen:
  Simulating {R}aman spectra using molecular dynamics},}\ }\href
  {http://link.aps.org/doi/10.1103/PhysRevB.87.174110} {\bibfield  {journal}
  {\bibinfo  {journal} {Phys. Rev. B}\ }\textbf {\bibinfo {volume} {87}},\
  \bibinfo {pages} {174110} (\bibinfo {year} {2013})}\BibitemShut {NoStop}%
\bibitem [{\citenamefont {Singh}\ \emph {et~al.}(2014)\citenamefont {Singh},
  \citenamefont {Azadi},\ and\ \citenamefont {K\"uhne}}]{md_ramand_ir_azadi}%
  \BibitemOpen
  \bibfield  {author} {\bibinfo {author} {\bibfnamefont {Ranber}\ \bibnamefont
  {Singh}}, \bibinfo {author} {\bibfnamefont {Sam}\ \bibnamefont {Azadi}}, \
  and\ \bibinfo {author} {\bibfnamefont {Thomas~D.}\ \bibnamefont {K\"uhne}},\
  }\bibfield  {title} {\enquote {\bibinfo {title} {Anharmonicity and
  finite-temperature effects on the structure, stability, and vibrational
  spectrum of phase {III} of solid molecular hydrogen},}\ }\href
  {http://link.aps.org/doi/10.1103/PhysRevB.90.014110} {\bibfield  {journal}
  {\bibinfo  {journal} {Phys. Rev. B}\ }\textbf {\bibinfo {volume} {90}},\
  \bibinfo {pages} {014110} (\bibinfo {year} {2014})}\BibitemShut {NoStop}%
\bibitem [{\citenamefont {Chen}\ \emph {et~al.}(2014)\citenamefont {Chen},
  \citenamefont {Ren}, \citenamefont {Li}, \citenamefont {Alf\`e},\ and\
  \citenamefont {Wang}}]{Chen_MD_DMC_2014}%
  \BibitemOpen
  \bibfield  {author} {\bibinfo {author} {\bibfnamefont {Ji}~\bibnamefont
  {Chen}}, \bibinfo {author} {\bibfnamefont {Xinguo}\ \bibnamefont {Ren}},
  \bibinfo {author} {\bibfnamefont {Xin-Zheng}\ \bibnamefont {Li}}, \bibinfo
  {author} {\bibfnamefont {Dario}\ \bibnamefont {Alf\`e}}, \ and\ \bibinfo
  {author} {\bibfnamefont {Enge}\ \bibnamefont {Wang}},\ }\bibfield  {title}
  {\enquote {\bibinfo {title} {On the room-temperature phase diagram of high
  pressure hydrogen: {A}n {\it ab initio} molecular dynamics perspective and a
  diffusion {M}onte {C}arlo study},}\ }\href
  {http://scitation.aip.org/content/aip/journal/jcp/141/2/10.1063/1.4886075}
  {\bibfield  {journal} {\bibinfo  {journal} {J. Chem. Phys.}\ }\textbf
  {\bibinfo {volume} {141}},\ \bibinfo {pages} {024501} (\bibinfo {year}
  {2014})}\BibitemShut {NoStop}%
\bibitem [{\citenamefont {Kitamura}\ \emph {et~al.}(2000)\citenamefont
  {Kitamura}, \citenamefont {Tsuneyuki}, \citenamefont {Ogitsu},\ and\
  \citenamefont {Miyake}}]{Kitamura_H_PIMD_2000}%
  \BibitemOpen
  \bibfield  {author} {\bibinfo {author} {\bibfnamefont {Hikaru}\ \bibnamefont
  {Kitamura}}, \bibinfo {author} {\bibfnamefont {Shinji}\ \bibnamefont
  {Tsuneyuki}}, \bibinfo {author} {\bibfnamefont {Tadashi}\ \bibnamefont
  {Ogitsu}}, \ and\ \bibinfo {author} {\bibfnamefont {Takashi}\ \bibnamefont
  {Miyake}},\ }\bibfield  {title} {\enquote {\bibinfo {title} {Quantum
  distribution of protons in solid molecular hydrogen at megabar pressures},}\
  }\href {http://dx.doi.org/10.1038/35005027} {\bibfield  {journal} {\bibinfo
  {journal} {Nature}\ }\textbf {\bibinfo {volume} {404}},\ \bibinfo {pages}
  {259} (\bibinfo {year} {2000})}\BibitemShut {NoStop}%
\bibitem [{\citenamefont {Chen}\ \emph {et~al.}(2013)\citenamefont {Chen},
  \citenamefont {Li}, \citenamefont {Zhang}, \citenamefont {Probert},
  \citenamefont {Pickard}, \citenamefont {Needs}, \citenamefont {Michaelides},\
  and\ \citenamefont {Wang}}]{pimd_hydrogen}%
  \BibitemOpen
  \bibfield  {author} {\bibinfo {author} {\bibfnamefont {Ji}~\bibnamefont
  {Chen}}, \bibinfo {author} {\bibfnamefont {Xin-Zheng}\ \bibnamefont {Li}},
  \bibinfo {author} {\bibfnamefont {Qianfan}\ \bibnamefont {Zhang}}, \bibinfo
  {author} {\bibfnamefont {Matthew I.~J.}\ \bibnamefont {Probert}}, \bibinfo
  {author} {\bibfnamefont {Chris~J.}\ \bibnamefont {Pickard}}, \bibinfo
  {author} {\bibfnamefont {Richard~J.}\ \bibnamefont {Needs}}, \bibinfo
  {author} {\bibfnamefont {Angelos}\ \bibnamefont {Michaelides}}, \ and\
  \bibinfo {author} {\bibfnamefont {Enge}\ \bibnamefont {Wang}},\ }\bibfield
  {title} {\enquote {\bibinfo {title} {Quantum simulation of low-temperature
  metallic liquid hydrogen},}\ }\href {http://dx.doi.org/10.1038/ncomms3064}
  {\bibfield  {journal} {\bibinfo  {journal} {Nat. Commun.}\ }\textbf {\bibinfo
  {volume} {4}},\ \bibinfo {pages} {2064} (\bibinfo {year} {2013})}\BibitemShut
  {NoStop}%
\bibitem [{\citenamefont {Ceperley}\ and\ \citenamefont
  {Alder}(1987)}]{Ceperley_1987_H}%
  \BibitemOpen
  \bibfield  {author} {\bibinfo {author} {\bibfnamefont {D.~M.}\ \bibnamefont
  {Ceperley}}\ and\ \bibinfo {author} {\bibfnamefont {B.~J.}\ \bibnamefont
  {Alder}},\ }\bibfield  {title} {\enquote {\bibinfo {title} {Ground state of
  solid hydrogen at high pressures},}\ }\href
  {http://link.aps.org/doi/10.1103/PhysRevB.36.2092} {\bibfield  {journal}
  {\bibinfo  {journal} {Phys. Rev. B}\ }\textbf {\bibinfo {volume} {36}},\
  \bibinfo {pages} {2092--2106} (\bibinfo {year} {1987})}\BibitemShut {NoStop}%
\bibitem [{\citenamefont {Azadi}\ \emph {et~al.}(2013)\citenamefont {Azadi},
  \citenamefont {Foulkes},\ and\ \citenamefont
  {K\"{u}hne}}]{azadi_hydrogen_gw_gap}%
  \BibitemOpen
  \bibfield  {author} {\bibinfo {author} {\bibfnamefont {Sam}\ \bibnamefont
  {Azadi}}, \bibinfo {author} {\bibfnamefont {W.~M.~C.}\ \bibnamefont
  {Foulkes}}, \ and\ \bibinfo {author} {\bibfnamefont {Thomas~D.}\ \bibnamefont
  {K\"{u}hne}},\ }\bibfield  {title} {\enquote {\bibinfo {title} {Quantum
  {M}onte {C}arlo study of high pressure solid molecular hydrogen},}\ }\href
  {http://stacks.iop.org/1367-2630/15/i=11/a=113005} {\bibfield  {journal}
  {\bibinfo  {journal} {New J. Phys.}\ }\textbf {\bibinfo {volume} {15}},\
  \bibinfo {pages} {113005} (\bibinfo {year} {2013})}\BibitemShut {NoStop}%
\bibitem [{\citenamefont {Azadi}\ \emph {et~al.}(2014)\citenamefont {Azadi},
  \citenamefont {Monserrat}, \citenamefont {Foulkes},\ and\ \citenamefont
  {Needs}}]{prl_dissociation_hydrogen}%
  \BibitemOpen
  \bibfield  {author} {\bibinfo {author} {\bibfnamefont {Sam}\ \bibnamefont
  {Azadi}}, \bibinfo {author} {\bibfnamefont {Bartomeu}\ \bibnamefont
  {Monserrat}}, \bibinfo {author} {\bibfnamefont {W.~M.~C.}\ \bibnamefont
  {Foulkes}}, \ and\ \bibinfo {author} {\bibfnamefont {R.~J.}\ \bibnamefont
  {Needs}},\ }\bibfield  {title} {\enquote {\bibinfo {title} {Dissociation of
  high-pressure solid molecular hydrogen: A quantum {M}onte {C}arlo and
  anharmonic vibrational study},}\ }\href
  {http://link.aps.org/doi/10.1103/PhysRevLett.112.165501} {\bibfield
  {journal} {\bibinfo  {journal} {Phys. Rev. Lett.}\ }\textbf {\bibinfo
  {volume} {112}},\ \bibinfo {pages} {165501} (\bibinfo {year}
  {2014})}\BibitemShut {NoStop}%
\bibitem [{\citenamefont {McMinis}\ \emph {et~al.}(2015)\citenamefont
  {McMinis}, \citenamefont {Clay}, \citenamefont {Lee},\ and\ \citenamefont
  {Morales}}]{h_dissociation_morales}%
  \BibitemOpen
  \bibfield  {author} {\bibinfo {author} {\bibfnamefont {Jeremy}\ \bibnamefont
  {McMinis}}, \bibinfo {author} {\bibfnamefont {Raymond~C.}\ \bibnamefont
  {Clay}}, \bibinfo {author} {\bibfnamefont {Donghwa}\ \bibnamefont {Lee}}, \
  and\ \bibinfo {author} {\bibfnamefont {Miguel~A.}\ \bibnamefont {Morales}},\
  }\bibfield  {title} {\enquote {\bibinfo {title} {Molecular to atomic phase
  transition in hydrogen under high pressure},}\ }\href
  {http://link.aps.org/doi/10.1103/PhysRevLett.114.105305} {\bibfield
  {journal} {\bibinfo  {journal} {Phys. Rev. Lett.}\ }\textbf {\bibinfo
  {volume} {114}},\ \bibinfo {pages} {105305} (\bibinfo {year}
  {2015})}\BibitemShut {NoStop}%
\bibitem [{\citenamefont {Drummond}\ \emph {et~al.}(2015)\citenamefont
  {Drummond}, \citenamefont {Monserrat}, \citenamefont {Lloyd-Williams},
  \citenamefont {L\'{o}pez~R\'{i}os}, \citenamefont {Pickard},\ and\
  \citenamefont {Needs}}]{hydrogen_nature_communications}%
  \BibitemOpen
  \bibfield  {author} {\bibinfo {author} {\bibfnamefont {N.~D.}\ \bibnamefont
  {Drummond}}, \bibinfo {author} {\bibfnamefont {B.}~\bibnamefont {Monserrat}},
  \bibinfo {author} {\bibfnamefont {J.~H.}\ \bibnamefont {Lloyd-Williams}},
  \bibinfo {author} {\bibfnamefont {P.}~\bibnamefont {L\'{o}pez~R\'{i}os}},
  \bibinfo {author} {\bibfnamefont {C.~J.}\ \bibnamefont {Pickard}}, \ and\
  \bibinfo {author} {\bibfnamefont {R.~J.}\ \bibnamefont {Needs}},\ }\bibfield
  {title} {\enquote {\bibinfo {title} {Quantum {M}onte {C}arlo study of the
  phase diagram of solid molecular hydrogen at extreme pressures},}\ }\href
  {http://dx.doi.org/10.1038/ncomms8794} {\bibfield  {journal} {\bibinfo
  {journal} {Nat. Commun.}\ }\textbf {\bibinfo {volume} {6}},\ \bibinfo {pages}
  {7794} (\bibinfo {year} {2015})}\BibitemShut {NoStop}%
\bibitem [{\citenamefont {Azadi}\ \emph {et~al.}(2016)\citenamefont {Azadi},
  \citenamefont {Drummond},\ and\ \citenamefont
  {Foulkes}}]{azadi_h_metal_dmc_static}%
  \BibitemOpen
  \bibfield  {author} {\bibinfo {author} {\bibfnamefont {Sam}\ \bibnamefont
  {Azadi}}, \bibinfo {author} {\bibfnamefont {N.~D.}\ \bibnamefont {Drummond}},
  \ and\ \bibinfo {author} {\bibfnamefont {W.~M.~C.}\ \bibnamefont {Foulkes}},\
  }\bibfield  {title} {\enquote {\bibinfo {title} {Nature of the metallization
  transition in solid hydrogen},}\ }\href {http://arxiv.org/abs/1608.00754}
  {\bibfield  {journal} {\bibinfo  {journal} {arXiv:1608.00754}\ } (\bibinfo
  {year} {2016})}\BibitemShut {NoStop}%
\bibitem [{\citenamefont {Johnson}\ and\ \citenamefont
  {Ashcroft}(2000)}]{Johnson_Aschroft_2000}%
  \BibitemOpen
  \bibfield  {author} {\bibinfo {author} {\bibfnamefont {Kurt~A.}\ \bibnamefont
  {Johnson}}\ and\ \bibinfo {author} {\bibfnamefont {N.~W.}\ \bibnamefont
  {Ashcroft}},\ }\bibfield  {title} {\enquote {\bibinfo {title} {Structure and
  bandgap closure in dense hydrogen},}\ }\href
  {http://dx.doi.org/10.1038/35001024} {\bibfield  {journal} {\bibinfo
  {journal} {Nature}\ }\textbf {\bibinfo {volume} {403}},\ \bibinfo {pages}
  {632} (\bibinfo {year} {2000})}\BibitemShut {NoStop}%
\bibitem [{\citenamefont {Pickard}\ and\ \citenamefont
  {Needs}(2007{\natexlab{a}})}]{nature_physics_h}%
  \BibitemOpen
  \bibfield  {author} {\bibinfo {author} {\bibfnamefont {Chris~J.}\
  \bibnamefont {Pickard}}\ and\ \bibinfo {author} {\bibfnamefont {Richard~J.}\
  \bibnamefont {Needs}},\ }\bibfield  {title} {\enquote {\bibinfo {title}
  {Structure of phase {III} of solid hydrogen},}\ }\href
  {http://dx.doi.org/10.1038/nphys625} {\bibfield  {journal} {\bibinfo
  {journal} {Nat. Phys.}\ }\textbf {\bibinfo {volume} {3}},\ \bibinfo {pages}
  {473--476} (\bibinfo {year} {2007}{\natexlab{a}})}\BibitemShut {NoStop}%
\bibitem [{\citenamefont {Pickard}\ \emph
  {et~al.}(2012{\natexlab{a}})\citenamefont {Pickard}, \citenamefont
  {Martinez-Canales},\ and\ \citenamefont {Needs}}]{phase_iv_prb}%
  \BibitemOpen
  \bibfield  {author} {\bibinfo {author} {\bibfnamefont {Chris~J.}\
  \bibnamefont {Pickard}}, \bibinfo {author} {\bibfnamefont {Miguel}\
  \bibnamefont {Martinez-Canales}}, \ and\ \bibinfo {author} {\bibfnamefont
  {Richard~J.}\ \bibnamefont {Needs}},\ }\bibfield  {title} {\enquote {\bibinfo
  {title} {Density functional theory study of phase {IV} of solid hydrogen},}\
  }\href {http://link.aps.org/doi/10.1103/PhysRevB.85.214114} {\bibfield
  {journal} {\bibinfo  {journal} {Phys. Rev. B}\ }\textbf {\bibinfo {volume}
  {85}},\ \bibinfo {pages} {214114} (\bibinfo {year}
  {2012}{\natexlab{a}})}\BibitemShut {NoStop}%
\bibitem [{\citenamefont {Pickard}\ \emph
  {et~al.}(2012{\natexlab{b}})\citenamefont {Pickard}, \citenamefont
  {Martinez-Canales},\ and\ \citenamefont {Needs}}]{phase_iv_prb_erratum}%
  \BibitemOpen
  \bibfield  {author} {\bibinfo {author} {\bibfnamefont {Chris~J.}\
  \bibnamefont {Pickard}}, \bibinfo {author} {\bibfnamefont {Miguel}\
  \bibnamefont {Martinez-Canales}}, \ and\ \bibinfo {author} {\bibfnamefont
  {Richard~J.}\ \bibnamefont {Needs}},\ }\bibfield  {title} {\enquote {\bibinfo
  {title} {Erratum: Density functional theory study of phase {IV} of solid
  hydrogen [{P}hys. {R}ev. {B} \textbf{85}, 214114 (2012)]},}\ }\href
  {http://link.aps.org/doi/10.1103/PhysRevB.86.059902} {\bibfield  {journal}
  {\bibinfo  {journal} {Phys. Rev. B}\ }\textbf {\bibinfo {volume} {86}},\
  \bibinfo {pages} {059902(E)} (\bibinfo {year}
  {2012}{\natexlab{b}})}\BibitemShut {NoStop}%
\bibitem [{\citenamefont {Azadi}\ and\ \citenamefont
  {K{\"u}hne}(2012)}]{azadi_hybrid_functionals}%
  \BibitemOpen
  \bibfield  {author} {\bibinfo {author} {\bibfnamefont {S.}~\bibnamefont
  {Azadi}}\ and\ \bibinfo {author} {\bibfnamefont {Th.~D.}\ \bibnamefont
  {K{\"u}hne}},\ }\bibfield  {title} {\enquote {\bibinfo {title} {Absence of
  metallization in solid molecular hydrogen},}\ }\href
  {http://dx.doi.org/10.1134/S0021364012090020} {\bibfield  {journal} {\bibinfo
   {journal} {JETP Lett.}\ }\textbf {\bibinfo {volume} {95}},\ \bibinfo {pages}
  {449} (\bibinfo {year} {2012})}\BibitemShut {NoStop}%
\bibitem [{\citenamefont {Dvorak}\ \emph {et~al.}(2014)\citenamefont {Dvorak},
  \citenamefont {Chen},\ and\ \citenamefont
  {Wu}}]{qp_excitonic_hydrogen_gw_bse}%
  \BibitemOpen
  \bibfield  {author} {\bibinfo {author} {\bibfnamefont {Marc}\ \bibnamefont
  {Dvorak}}, \bibinfo {author} {\bibfnamefont {Xiao-Jia}\ \bibnamefont {Chen}},
  \ and\ \bibinfo {author} {\bibfnamefont {Zhigang}\ \bibnamefont {Wu}},\
  }\bibfield  {title} {\enquote {\bibinfo {title} {Quasiparticle energies and
  excitonic effects in dense solid hydrogen near metallization},}\ }\href
  {\doibase 10.1103/PhysRevB.90.035103} {\bibfield  {journal} {\bibinfo
  {journal} {Phys. Rev. B}\ }\textbf {\bibinfo {volume} {90}},\ \bibinfo
  {pages} {035103} (\bibinfo {year} {2014})}\BibitemShut {NoStop}%
\bibitem [{\citenamefont {Pickard}\ and\ \citenamefont
  {Needs}(2009)}]{Pickard_PSS_2009}%
  \BibitemOpen
  \bibfield  {author} {\bibinfo {author} {\bibfnamefont {C.~J.}\ \bibnamefont
  {Pickard}}\ and\ \bibinfo {author} {\bibfnamefont {R.~J.}\ \bibnamefont
  {Needs}},\ }\bibfield  {title} {\enquote {\bibinfo {title} {Structures at
  high pressure from random searching},}\ }\href {\doibase
  10.1002/pssb.200880546} {\bibfield  {journal} {\bibinfo  {journal} {Phys.
  Status Solidi B}\ }\textbf {\bibinfo {volume} {246}},\ \bibinfo {pages}
  {536--540} (\bibinfo {year} {2009})}\BibitemShut {NoStop}%
\bibitem [{\citenamefont {Pickard}\ and\ \citenamefont
  {Needs}(2006)}]{PhysRevLett.97.045504}%
  \BibitemOpen
  \bibfield  {author} {\bibinfo {author} {\bibfnamefont {Chris~J.}\
  \bibnamefont {Pickard}}\ and\ \bibinfo {author} {\bibfnamefont {R.~J.}\
  \bibnamefont {Needs}},\ }\bibfield  {title} {\enquote {\bibinfo {title}
  {High-pressure phases of silane},}\ }\href
  {http://link.aps.org/doi/10.1103/PhysRevLett.97.045504} {\bibfield  {journal}
  {\bibinfo  {journal} {Phys. Rev. Lett.}\ }\textbf {\bibinfo {volume} {97}},\
  \bibinfo {pages} {045504} (\bibinfo {year} {2006})}\BibitemShut {NoStop}%
\bibitem [{\citenamefont {Pickard}\ and\ \citenamefont
  {Needs}(2007{\natexlab{b}})}]{Aluminum_hydride_2007}%
  \BibitemOpen
  \bibfield  {author} {\bibinfo {author} {\bibfnamefont {Chris~J.}\
  \bibnamefont {Pickard}}\ and\ \bibinfo {author} {\bibfnamefont {R.~J.}\
  \bibnamefont {Needs}},\ }\bibfield  {title} {\enquote {\bibinfo {title}
  {Metallization of aluminum hydride at high pressures: A first-principles
  study},}\ }\href {http://link.aps.org/doi/10.1103/PhysRevB.76.144114}
  {\bibfield  {journal} {\bibinfo  {journal} {Phys. Rev. B}\ }\textbf {\bibinfo
  {volume} {76}},\ \bibinfo {pages} {144114} (\bibinfo {year}
  {2007}{\natexlab{b}})}\BibitemShut {NoStop}%
\bibitem [{\citenamefont {Pickard}\ and\ \citenamefont
  {Needs}(2008)}]{Ammonia_2008}%
  \BibitemOpen
  \bibfield  {author} {\bibinfo {author} {\bibfnamefont {Chris~J.}\
  \bibnamefont {Pickard}}\ and\ \bibinfo {author} {\bibfnamefont {R.~J.}\
  \bibnamefont {Needs}},\ }\bibfield  {title} {\enquote {\bibinfo {title}
  {Highly compressed ammonia forms an ionic crystal},}\ }\href
  {http://dx.doi.org/10.1038/nmat2261} {\bibfield  {journal} {\bibinfo
  {journal} {Nat. Mater.}\ }\textbf {\bibinfo {volume} {7}},\ \bibinfo {pages}
  {775} (\bibinfo {year} {2008})}\BibitemShut {NoStop}%
\bibitem [{\citenamefont {Ninet}\ \emph {et~al.}(2014)\citenamefont {Ninet},
  \citenamefont {Datchi}, \citenamefont {Dumas}, \citenamefont {Mezouar},
  \citenamefont {Garbarino}, \citenamefont {Mafety}, \citenamefont {Pickard},
  \citenamefont {Needs},\ and\ \citenamefont {Saitta}}]{Ammonia_2014}%
  \BibitemOpen
  \bibfield  {author} {\bibinfo {author} {\bibfnamefont {S.}~\bibnamefont
  {Ninet}}, \bibinfo {author} {\bibfnamefont {F.}~\bibnamefont {Datchi}},
  \bibinfo {author} {\bibfnamefont {P.}~\bibnamefont {Dumas}}, \bibinfo
  {author} {\bibfnamefont {M.}~\bibnamefont {Mezouar}}, \bibinfo {author}
  {\bibfnamefont {G.}~\bibnamefont {Garbarino}}, \bibinfo {author}
  {\bibfnamefont {A.}~\bibnamefont {Mafety}}, \bibinfo {author} {\bibfnamefont
  {C.~J.}\ \bibnamefont {Pickard}}, \bibinfo {author} {\bibfnamefont {R.~J.}\
  \bibnamefont {Needs}}, \ and\ \bibinfo {author} {\bibfnamefont {A.~M.}\
  \bibnamefont {Saitta}},\ }\bibfield  {title} {\enquote {\bibinfo {title}
  {Experimental and theoretical evidence for an ionic crystal of ammonia at
  high pressure},}\ }\href {http://link.aps.org/doi/10.1103/PhysRevB.89.174103}
  {\bibfield  {journal} {\bibinfo  {journal} {Phys. Rev. B}\ }\textbf {\bibinfo
  {volume} {89}},\ \bibinfo {pages} {174103} (\bibinfo {year}
  {2014})}\BibitemShut {NoStop}%
\bibitem [{\citenamefont {Fortes}\ \emph {et~al.}(2009)\citenamefont {Fortes},
  \citenamefont {Suard}, \citenamefont {Lem\'{e}e-Cailleau}, \citenamefont
  {Pickard},\ and\ \citenamefont {Needs}}]{ADMII_2009}%
  \BibitemOpen
  \bibfield  {author} {\bibinfo {author} {\bibfnamefont {A.~Dominic}\
  \bibnamefont {Fortes}}, \bibinfo {author} {\bibfnamefont {Emmanuelle}\
  \bibnamefont {Suard}}, \bibinfo {author} {\bibfnamefont
  {Marie-H\'{e}l\`{e}ne}\ \bibnamefont {Lem\'{e}e-Cailleau}}, \bibinfo {author}
  {\bibfnamefont {Christopher~J.}\ \bibnamefont {Pickard}}, \ and\ \bibinfo
  {author} {\bibfnamefont {Richard~J.}\ \bibnamefont {Needs}},\ }\bibfield
  {title} {\enquote {\bibinfo {title} {Crystal structure of ammonia monohydrate
  phase {II}},}\ }\href {http://dx.doi.org/10.1021/ja9052569} {\bibfield
  {journal} {\bibinfo  {journal} {J. Amer. Chem. Soc.}\ }\textbf {\bibinfo
  {volume} {131}},\ \bibinfo {pages} {13508--13515} (\bibinfo {year}
  {2009})}\BibitemShut {NoStop}%
\bibitem [{\citenamefont {Dewaele}\ \emph {et~al.}(2016)\citenamefont
  {Dewaele}, \citenamefont {Worth}, \citenamefont {Pickard}, \citenamefont
  {Needs}, \citenamefont {Pascarelli}, \citenamefont {Mathon}, \citenamefont
  {Mezouar},\ and\ \citenamefont {Irifune}}]{xenon_oxides}%
  \BibitemOpen
  \bibfield  {author} {\bibinfo {author} {\bibfnamefont {Agn\`es}\ \bibnamefont
  {Dewaele}}, \bibinfo {author} {\bibfnamefont {Nicholas}\ \bibnamefont
  {Worth}}, \bibinfo {author} {\bibfnamefont {Chris~J.}\ \bibnamefont
  {Pickard}}, \bibinfo {author} {\bibfnamefont {Richard~J.}\ \bibnamefont
  {Needs}}, \bibinfo {author} {\bibfnamefont {Sakura}\ \bibnamefont
  {Pascarelli}}, \bibinfo {author} {\bibfnamefont {Olivier}\ \bibnamefont
  {Mathon}}, \bibinfo {author} {\bibfnamefont {Mohamed}\ \bibnamefont
  {Mezouar}}, \ and\ \bibinfo {author} {\bibfnamefont {Tetsuo}\ \bibnamefont
  {Irifune}},\ }\bibfield  {title} {\enquote {\bibinfo {title} {Synthesis and
  stability of xenon oxides {X}e$_2${O}$_5$ and {X}e$_3${O}$_2$ under
  pressure},}\ }\href {http://dx.doi.org/10.1038/nchem.2528} {\bibfield
  {journal} {\bibinfo  {journal} {Nat. Chem.}\ }\textbf {\bibinfo {volume}
  {8}},\ \bibinfo {pages} {784} (\bibinfo {year} {2016})}\BibitemShut {NoStop}%
\bibitem [{\citenamefont {Goncharov}\ \emph {et~al.}(2013)\citenamefont
  {Goncharov}, \citenamefont {Howie},\ and\ \citenamefont
  {Gregoryanz}}]{h_exp_review}%
  \BibitemOpen
  \bibfield  {author} {\bibinfo {author} {\bibfnamefont {Alexander~F.}\
  \bibnamefont {Goncharov}}, \bibinfo {author} {\bibfnamefont {Ross~T.}\
  \bibnamefont {Howie}}, \ and\ \bibinfo {author} {\bibfnamefont {Eugene}\
  \bibnamefont {Gregoryanz}},\ }\bibfield  {title} {\enquote {\bibinfo {title}
  {Hydrogen at extreme pressures},}\ }\href
  {http://scitation.aip.org/content/aip/journal/ltp/39/5/10.1063/1.4807051}
  {\bibfield  {journal} {\bibinfo  {journal} {Low Temp. Phys.}\ }\textbf
  {\bibinfo {volume} {39}},\ \bibinfo {pages} {402--408} (\bibinfo {year}
  {2013})}\BibitemShut {NoStop}%
\bibitem [{\citenamefont {Jones}(2015)}]{Jones_review_DFT_2015}%
  \BibitemOpen
  \bibfield  {author} {\bibinfo {author} {\bibfnamefont {R.~O.}\ \bibnamefont
  {Jones}},\ }\bibfield  {title} {\enquote {\bibinfo {title} {Density
  functional theory: Its origins, rise to prominence, and future},}\ }\href
  {http://link.aps.org/doi/10.1103/RevModPhys.87.897} {\bibfield  {journal}
  {\bibinfo  {journal} {Rev. Mod. Phys.}\ }\textbf {\bibinfo {volume} {87}},\
  \bibinfo {pages} {897--923} (\bibinfo {year} {2015})}\BibitemShut {NoStop}%
\bibitem [{\citenamefont {Clark}\ \emph {et~al.}(2005)\citenamefont {Clark},
  \citenamefont {Segall}, \citenamefont {Pickard}, \citenamefont {Hasnip},
  \citenamefont {Probert}, \citenamefont {Refson},\ and\ \citenamefont
  {Payne}}]{CASTEP}%
  \BibitemOpen
  \bibfield  {author} {\bibinfo {author} {\bibfnamefont {Stewart~J.}\
  \bibnamefont {Clark}}, \bibinfo {author} {\bibfnamefont {Matthew~D.}\
  \bibnamefont {Segall}}, \bibinfo {author} {\bibfnamefont {Chris~J.}\
  \bibnamefont {Pickard}}, \bibinfo {author} {\bibfnamefont {Phil~J.}\
  \bibnamefont {Hasnip}}, \bibinfo {author} {\bibfnamefont {Matt I.~J.}\
  \bibnamefont {Probert}}, \bibinfo {author} {\bibfnamefont {Keith}\
  \bibnamefont {Refson}}, \ and\ \bibinfo {author} {\bibfnamefont {Mike~C.}\
  \bibnamefont {Payne}},\ }\bibfield  {title} {\enquote {\bibinfo {title}
  {First principles methods using {\sc castep}},}\ }\href
  {http://www.oldenbourg-link.com/doi/abs/10.1524/zkri.220.5.567.65075}
  {\bibfield  {journal} {\bibinfo  {journal} {Z. Kristallogr.}\ }\textbf
  {\bibinfo {volume} {220}},\ \bibinfo {pages} {567} (\bibinfo {year}
  {2005})}\BibitemShut {NoStop}%
\bibitem [{\citenamefont {Vanderbilt}(1990)}]{PhysRevB.41.7892}%
  \BibitemOpen
  \bibfield  {author} {\bibinfo {author} {\bibfnamefont {David}\ \bibnamefont
  {Vanderbilt}},\ }\bibfield  {title} {\enquote {\bibinfo {title} {Soft
  self-consistent pseudopotentials in a generalized eigenvalue formalism},}\
  }\href {http://link.aps.org/doi/10.1103/PhysRevB.41.7892} {\bibfield
  {journal} {\bibinfo  {journal} {Phys. Rev. B}\ }\textbf {\bibinfo {volume}
  {41}},\ \bibinfo {pages} {7892--7895} (\bibinfo {year} {1990})}\BibitemShut
  {NoStop}%
\bibitem [{\citenamefont {Becke}(1988)}]{blyp_exchange}%
  \BibitemOpen
  \bibfield  {author} {\bibinfo {author} {\bibfnamefont {A.~D.}\ \bibnamefont
  {Becke}},\ }\bibfield  {title} {\enquote {\bibinfo {title}
  {Density-functional exchange-energy approximation with correct asymptotic
  behavior},}\ }\href {http://link.aps.org/doi/10.1103/PhysRevA.38.3098}
  {\bibfield  {journal} {\bibinfo  {journal} {Phys. Rev. A}\ }\textbf {\bibinfo
  {volume} {38}},\ \bibinfo {pages} {3098--3100} (\bibinfo {year}
  {1988})}\BibitemShut {NoStop}%
\bibitem [{\citenamefont {Lee}\ \emph {et~al.}(1988)\citenamefont {Lee},
  \citenamefont {Yang},\ and\ \citenamefont {Parr}}]{blyp_correlation}%
  \BibitemOpen
  \bibfield  {author} {\bibinfo {author} {\bibfnamefont {Chengteh}\
  \bibnamefont {Lee}}, \bibinfo {author} {\bibfnamefont {Weitao}\ \bibnamefont
  {Yang}}, \ and\ \bibinfo {author} {\bibfnamefont {Robert~G.}\ \bibnamefont
  {Parr}},\ }\bibfield  {title} {\enquote {\bibinfo {title} {Development of the
  {C}olle-{S}alvetti correlation-energy formula into a functional of the
  electron density},}\ }\href {http://link.aps.org/doi/10.1103/PhysRevB.37.785}
  {\bibfield  {journal} {\bibinfo  {journal} {Phys. Rev. B}\ }\textbf {\bibinfo
  {volume} {37}},\ \bibinfo {pages} {785--789} (\bibinfo {year}
  {1988})}\BibitemShut {NoStop}%
\bibitem [{\citenamefont {Clay}\ \emph {et~al.}(2014)\citenamefont {Clay},
  \citenamefont {Mcminis}, \citenamefont {McMahon}, \citenamefont {Pierleoni},
  \citenamefont {Ceperley},\ and\ \citenamefont {Morales}}]{clay_benchmarking}%
  \BibitemOpen
  \bibfield  {author} {\bibinfo {author} {\bibfnamefont {Raymond~C.}\
  \bibnamefont {Clay}}, \bibinfo {author} {\bibfnamefont {Jeremy}\ \bibnamefont
  {Mcminis}}, \bibinfo {author} {\bibfnamefont {Jeffrey~M.}\ \bibnamefont
  {McMahon}}, \bibinfo {author} {\bibfnamefont {Carlo}\ \bibnamefont
  {Pierleoni}}, \bibinfo {author} {\bibfnamefont {David~M.}\ \bibnamefont
  {Ceperley}}, \ and\ \bibinfo {author} {\bibfnamefont {Miguel~A.}\
  \bibnamefont {Morales}},\ }\bibfield  {title} {\enquote {\bibinfo {title}
  {Benchmarking exchange-correlation functionals for hydrogen at high pressures
  using quantum {M}onte {C}arlo},}\ }\href
  {http://link.aps.org/doi/10.1103/PhysRevB.89.184106} {\bibfield  {journal}
  {\bibinfo  {journal} {Phys. Rev. B}\ }\textbf {\bibinfo {volume} {89}},\
  \bibinfo {pages} {184106} (\bibinfo {year} {2014})}\BibitemShut {NoStop}%
\bibitem [{\citenamefont {Monserrat}\ \emph {et~al.}(2013)\citenamefont
  {Monserrat}, \citenamefont {Drummond},\ and\ \citenamefont
  {Needs}}]{PhysRevB.87.144302}%
  \BibitemOpen
  \bibfield  {author} {\bibinfo {author} {\bibfnamefont {Bartomeu}\
  \bibnamefont {Monserrat}}, \bibinfo {author} {\bibfnamefont {N.~D.}\
  \bibnamefont {Drummond}}, \ and\ \bibinfo {author} {\bibfnamefont {R.~J.}\
  \bibnamefont {Needs}},\ }\bibfield  {title} {\enquote {\bibinfo {title}
  {Anharmonic vibrational properties in periodic systems: Energy,
  electron-phonon coupling, and stress},}\ }\href
  {http://link.aps.org/doi/10.1103/PhysRevB.87.144302} {\bibfield  {journal}
  {\bibinfo  {journal} {Phys. Rev. B}\ }\textbf {\bibinfo {volume} {87}},\
  \bibinfo {pages} {144302} (\bibinfo {year} {2013})}\BibitemShut {NoStop}%
\bibitem [{\citenamefont {Lloyd-Williams}\ and\ \citenamefont
  {Monserrat}(2015)}]{non_diagonal}%
  \BibitemOpen
  \bibfield  {author} {\bibinfo {author} {\bibfnamefont {Jonathan~H.}\
  \bibnamefont {Lloyd-Williams}}\ and\ \bibinfo {author} {\bibfnamefont
  {Bartomeu}\ \bibnamefont {Monserrat}},\ }\bibfield  {title} {\enquote
  {\bibinfo {title} {Lattice dynamics and electron-phonon coupling calculations
  using nondiagonal supercells},}\ }\href
  {http://link.aps.org/doi/10.1103/PhysRevB.92.184301} {\bibfield  {journal}
  {\bibinfo  {journal} {Phys. Rev. B}\ }\textbf {\bibinfo {volume} {92}},\
  \bibinfo {pages} {184301} (\bibinfo {year} {2015})}\BibitemShut {NoStop}%
\bibitem [{\citenamefont {Perdew}\ \emph {et~al.}(1996)\citenamefont {Perdew},
  \citenamefont {Burke},\ and\ \citenamefont
  {Ernzerhof}}]{PhysRevLett.77.3865}%
  \BibitemOpen
  \bibfield  {author} {\bibinfo {author} {\bibfnamefont {John~P.}\ \bibnamefont
  {Perdew}}, \bibinfo {author} {\bibfnamefont {Kieron}\ \bibnamefont {Burke}},
  \ and\ \bibinfo {author} {\bibfnamefont {Matthias}\ \bibnamefont
  {Ernzerhof}},\ }\bibfield  {title} {\enquote {\bibinfo {title} {Generalized
  gradient approximation made simple},}\ }\href
  {http://link.aps.org/doi/10.1103/PhysRevLett.77.3865} {\bibfield  {journal}
  {\bibinfo  {journal} {Phys. Rev. Lett.}\ }\textbf {\bibinfo {volume} {77}},\
  \bibinfo {pages} {3865} (\bibinfo {year} {1996})}\BibitemShut {NoStop}%
\end{thebibliography}%

\end{document}